\documentclass[prd,superscriptaddress,preprintnumbers,tightenlines,nofootinbib, eqsecnum]{revtex4-2}

\usepackage{amsmath}
\usepackage{amsfonts}
\usepackage{amssymb}
\usepackage{bm}
\usepackage{hyperref}
\usepackage{mathrsfs}
\usepackage{graphicx}
\usepackage{empheq}
\usepackage{ulem}
\usepackage[usenames]{color}
\usepackage{ytableau}
        
\allowdisplaybreaks

\DeclareSymbolFontAlphabet{\mathrsfs}{rsfs}
\DeclareMathAlphabet{\mathcal}{OMS}{cmsy}{m}{n}

\newcommand{\FP}{\mathop{\mathrm{FP}}_{B=0}}

\newcommand{\BoxRm}{\,\overline{\Box}_{\mathrm{ret}}^{-1}}
\newcommand{\nn}{\nonumber}
\newcommand\calO{{\mathcal{O}}}
\newcommand{\dd}{\mathrm{d}}
\newcommand{\di}{\mathrm{i}} 
\newcommand{\de}{\mathrm{e}} 
\newcommand{\dI}{\mathrm{I}}
\newcommand{\dJ}{\mathrm{J}}
\newcommand{\dM}{\mathrm{M}}

\newcommand{\dS}{\mathrm{S}}

\newcommand{\dW}{\mathrm{W}}
\newcommand{\dX}{\mathrm{X}}
\newcommand{\dY}{\mathrm{Y}}
\newcommand{\dZ}{\mathrm{Z}}

\newcommand{\comm}{\mathcal{C}}

\definecolor{darkgreen}{rgb}{0,0.5,0}

\hypersetup{
 unicode=false,          % non-Latin characters in Acrobat???s bookmarks
 pdftoolbar=true,        % show Acrobat???s toolbar?
 pdfmenubar=true,        % show Acrobat???s menu?
 pdffitwindow=false,     % window fit to page when opened
 pdfstartview={FitH},    % fits the width of the page to the window
 pdftitle={My title},    % title
 pdfauthor={Author},     % author
 pdfsubject={Subject},   % subject of the document
 pdfcreator={Creator},   % creator of the document
 pdfproducer={Producer}, % producer of the document
 pdfkeywords={keyword1} {key2} {key3}, % list of keywords
 pdfnewwindow=true,      % links in new window
 colorlinks=true,       % false: boxed links; true: colored links
 linkcolor=red,          % color of internal links
 citecolor=cyan,        % color of links to bibliography
 filecolor=magenta,      % color of file links
 urlcolor=darkgreen,           % color of external links
 linktocpage=true
}

\makeatletter
\g@addto@macro\bfseries{\boldmath}
\makeatother

\begin{document}
	
\title{Gravitational Wave Flux and Quadrupole Modes from Quasi-Circular\\ Non-Spinning Compact Binaries to the Fourth Post-Newtonian Order}

\author{Luc \textsc{Blanchet}}\email{luc.blanchet@iap.fr}
\affiliation{$\mathcal{G}\mathbb{R}\varepsilon{\mathbb{C}}\mathcal{O}$, 
	Institut d'Astrophysique de Paris, \\UMR 7095, CNRS, Sorbonne Universit{\'e},
	98\textsuperscript{bis} boulevard Arago, 75014 Paris, France}

\author{Guillaume \textsc{Faye}}\email{faye@iap.fr}
\affiliation{$\mathcal{G}\mathbb{R}\varepsilon{\mathbb{C}}\mathcal{O}$, 
Institut d'Astrophysique de Paris, \\UMR 7095, CNRS, Sorbonne Universit{\'e},
98\textsuperscript{bis} boulevard Arago, 75014 Paris, France}
\affiliation{Centre for Strings, Gravitation and Cosmology, Department of Physics,
Indian Institute of Technology Madras, Chennai 600036, India}

\author{Quentin \textsc{Henry}}\email{quentin.henry@aei.mpg.de}
\affiliation{Max Planck Institute for Gravitational Physics\\
(Albert Einstein Institute), D-14476 Potsdam, Germany}

\author{Fran\c{c}ois \textsc{Larrouturou}}\email{francois.larrouturou@desy.de}
\affiliation{Deutsches Elektronen-Synchrotron DESY,\\
Notkestr. 85, 22607 Hamburg, Germany}

\author{David \textsc{Trestini}}\email{david.trestini@obspm.fr}
\affiliation{$\mathcal{G}\mathbb{R}\varepsilon{\mathbb{C}}\mathcal{O}$, 
	Institut d'Astrophysique de Paris, \\UMR 7095, CNRS, Sorbonne Universit{\'e},
	98\textsuperscript{bis} boulevard Arago, 75014 Paris, France}
	\affiliation{Laboratoire Univers et Th\'eories, Observatoire de Paris, Universit\'e PSL, Universit\'e Paris Cit\'e, CNRS, F-92190 Meudon, France}

\date{\today}

\begin{abstract}
This article provides the details on the technical derivation of the gravitational waveform and total gravitational-wave energy flux of non-spinning compact binary systems to the 4PN (fourth post-Newtonian) order beyond the Einstein quadrupole formula. In particular: (i) we overview the link between the radiative multipole moments measured at infinity and the source moments in the framework of dimensional regularization; (ii) we compute special corrections to the source moments due to ``infrared'' commutators arising at the 4PN order; (iii) we derive a ``post-adiabatic'' correction needed to evaluate the tail integral with 2.5PN relative precision; (iv) we discuss the relation between the binary's orbital frequency in quasi-circular orbit and the gravitational-wave frequency measured at infinity; (v) we compute the hereditary effects at the 4PN order, including those coming from the recently derived tails-of-memory; and (vi) we describe the various tests we have performed to ensure the correctness of the results. Those results are collected in an ancillary file.   
\end{abstract}

\pacs{04.25.Nx, 04.30.-w, 97.60.Jd, 97.60.Lf}

\preprint{DESY-23-044}

\maketitle

\section{Introduction}\label{sec:introduction}

The post-Newtonian (PN) approximation is a paramount tool for computing the generation of gravitational waves (GWs) by isolated sources. This perturbation technique relies on two joint expansions, valid for systems that have slow velocity and are self-gravitating, and which are thus linked by weak gravitational fields. In consequence, it is ideally suited to study the inspiral phase of compact binary systems, made of black holes (BHs) or neutron stars (NSs). While the derivation of lowest-order PN results, starting from the famous Einstein quadrupole formula~\cite{E18,LL,PM63}, is fairly straightforward, the systematic expansion to high PN orders entails a number of subtle issues. One is the control of physical non-linear effects in the propagation of GWs from source to detector. Other more technical difficulties are linked to the appearance of divergent integrals in the calculations, and the subsequent need to implement a proper regularization scheme. 

There are two modern ways of carrying out the PN approximation. The first one is to embed it into the more general PN-MPM framework~\cite{BD86,B87,BD88,BD92,B98mult}, consisting of a multipolar and post-Minkowskian (MPM) expansion for the field exterior to the isolated source, subsequently matched to the inner PN field of the source. This method achieved the completion of the gravitational waveform to 3.5PN order for non-spinning objects~\cite{BDIWW95,B98tail,BIJ02,BFIJ02,BI04mult,BDEI04,BDEI05dr,BFIS08,FMBI12,Henry:2022ccf}, \emph{i.e.}, up to the $(v/c)^7$ correction beyond the Einstein quadrupole formula. A variant of the PN-MPM framework is the DIRE formalism, developed to 2PN order~\cite{WW96}. The second way relies on an effective field theory (EFT)~\cite{GR06,GRoss10}, extracting the radiation sector from a derivative expansion at the level of the action~\cite{Ross:2012fc}. This more recent framework has reached the 2PN precision~\cite{LMRY19}. We refer the interested reader to the reviews~\cite{BlanchetLR,Maggiore,BuonSathya15,Porto16} for more details on the PN scheme.

The main output of the PN calculation is the observable waveform (phase and amplitude modes) as an expansion series in terms of the PN parameter $x=(\frac{G m \Omega}{c^3})^{2/3}$, where $\Omega$ is the half of the GW frequency. This is obviously essential for the data analysis of GW detectors~\cite{Cutler:1992tc,Cutler:1994ys}, notably for the current LIGO-Virgo-Kagra network~\cite{LIGOScientific:2018mvr,LIGOScientific:2020ibl,LIGOScientific:2021djp}, but also for future generations of detectors, such as the Laser Interferometer Space Antenna (LISA)~\cite{LISA:2017pwj} or the Einstein Telescope (ET)~\cite{ET:2022}. In addition, the PN results are of great interest for numerical relativity, as they are naturally used for comparisons with numerical results (see \emph{e.g.}~\cite{Grandclement:2007sb,Boyle:2007ft,Bernuzzi:2011aq,Borhanian:2019kxt}), as well as for the calibration of initial data (see \emph{e.g.}~\cite{Tichy:2002ec,Reifenberger:2012yg,Johnson-McDaniel:2009tvj,Healy:2017zqj}). They also play a crucial role for comparisons with gravitational self-force (GSF) results, notably second-order ones~\cite{Wardell:2021fyy,Albertini:2022rfe,Albertini:2022dmc}.

On the other hand, the PN framework is essential when it comes to experimentally testing the theory of gravitation. By drawing precise predictions and confronting them against observations, it allows performing ``agnostic'' tests of general relativity (GR) in the Solar system, for binary pulsars and even for transiting exoplanets~\cite{Blanchet:2019zxv}. With GWs, the idea is to treat each coefficient in the PN expansion of the GW phase as an independent parameter, and fit it against GW data~\cite{Bsat95,AIQS06a}. This method has been applied to data collected from LIGO-Virgo detectors, and has already permitted probing the non-linear structure of GR by confirming the signature of GW tails to the 1.5PN order~\cite{LIGOtestGR,LIGOtestGR2,LIGOScientific:2021sio}. This \emph{modus operandi} is very promising, notably regarding the possibility of multi-band detections, combining LISA and detectors on ground~\cite{Sesana16}.
In the same spirit, the prediction for the tidal effects affecting the binary NSs~\cite{Flanagan:2007ix, Damour:2009vw, VHF11, Damour:2012yf, BiniDF12, HFB20a, HFB20b, HFB20c}, which first contribute to the waveform at the 5PN order (in the spin-less case), 
%in the adiabatic spinless  approximation \cite{Damour:2009vw, Damour:2012yf, BiniDF12,HFB20a,HFB20b,HFB20c} (see also \cite{Steinhoff:2016rfi, Hinderer:2016eia, AGP18} going beyond that model),
%In the same spirit, the prediction for the tidal effects affecting binary NSs~\cite{BiniDF12,AGP18,HFB20a,HFB20b,HFB20c},which first contribute to the waveform at the 5PN order,
have been used at leading order to put a constraint on the equation of state of matter composing neutron stars~\cite{GW170817}, and could be used to distinguish usual BHs from ``exotic'' compact objects with ET. In addition to the agnostic tests, the PN framework is also used to perform ``educated'' tests, by computing observable quantities in specific alternative theories. It has been developed, for instance, in the context of massless scalar field theories~\cite{MW13,Laura1,Laura2,Bernard:2019yfz}, and used to make predictions concerning the gravitational radiation~\cite{Lang:2013fna,Lang:2014osa,Bernard:2022noq}. Similar studies have been conducted for various alternative theories~\cite{Sotiriou:2006pq,DeFelice:2010aj,Yagi:2011xp,Julie:2018lfp,Shiralilou:2021mfl}.

This paper presents the last step of the completion of the gravitational waveform for non-spinning, compact binary systems evolving with 4PN precision, in the case of quasi-circular orbits (and in GR). The first step was the computation of the equations of motion to 4PN order, from which one deduces in particular the binary's gravitational binding energy. The equations of motion were obtained by several groups using different methods: (i) the Arnowitt-Deser-Misner (ADM) Hamiltonian formalism~\cite{JaraS12, JaraS13, DJS14, DJS16} led to complete results except for one ``ambiguity'' parameter, which was fixed by resorting to a comparison with gravitational self-force results~\cite{BiniD13}; (ii) the Fokker Lagrangian formalism in harmonic coordinates~\cite{BBBFMa, BBBFMb,BBBFMc,BBFM17} yielded for the first time a complete result without ambiguity parameter; and (iii) the effective field theory (EFT) approach~\cite{FS4PN,FStail,GLPR16,FMSS16,PR17,FS19,FPRS19,Blumlein20} recovered the complete and unambiguous result. The second step was the computation and proper regularization, using the PN-MPM formalism, of the source mass quadrupole moment to 4PN~\cite{MHLMFB20,MQ4PN_IR,MQ4PN_renorm}, the current quadrupole to 3PN~\cite{HFB_courant} and the mass octupole to the same order~\cite{FBI15}. The next step, relying on previous works as well, was the control of the various non-linear interactions between moments occuring through the 4PN order~\cite{BFIS08,FMBI12,FBI15,MBF16,MQ4PN_jauge,TLB22,TB23}. The prominent of these interactions at 4PN order is the so-called ``tail-of-memory'', which is a cubically non-linear effect~\cite{TB23}. Moreover, we were able to compute the flux for circular orbits up to the 4.5PN order, since only one non-linear interaction plays a role for this computation, which is the so-called ``tail-of-tail-of-tail'', a quartic effect~\cite{MBF16} (see also~\cite{Messina:2017yjg}). In the present work, we complete the derivation of the gravitational flux to 4PN for generic orbits, and extract the $(\ell,\text{m}) = (2,2)$ and  $(\ell,\text{m}) = (2,0)$ modes from the radiative mass quadrupole moment. With those results at hand, and the 4.5PN term of the flux for quasi-circular orbits already computed in~\cite{MBF16}, the gravitational phasing at the 4.5PN precision is derived and reported in the companion Letter~\cite{BFHLT_Letter}.

This work is organized as follows. In Sec.~\ref{sec:nonlin}, we explicitly present the relations between the radiative, canonical and source multipole moments, at the required 4PN accuracy (and 4.5PN for circular orbits). Notably, we derive the missing relation between the canonical and source quadrupole moment, extending~\cite{MQ4PN_jauge} to arbitrary dimensions as required by dimensional regularization. Sec.~\ref{sec:commutators} is devoted to the calculation of novel terms coined as ``infrared (IR) commutators'' appearing at the 4PN order in the metric and the source mass quadrupole moment. They are zero for circular orbits and do not contribute to the results reported in~\cite{BFHLT_Letter}. Then, Sec.~\ref{sec:source} lists the expressions of the required source moments on circular orbits, obtained in previous works, notably~\cite{MHLMFB20,MQ4PN_IR,MQ4PN_renorm,HFB_courant,FBI15}. Sec.~\ref{sec:toolbox} exposes the technical toolbox that we used to perform the derivations and integrations needed to get the final results. Notably, an important physical contribution at 4PN order comes from the 1.5PN tail term computed to relative 2.5PN order, which necessitates a ``post-adiabatic'' evaluation, described in Sec.~\ref{sec:PA}. The results are presented in Sec.~\ref{sec:res}, and consist of: (i) the flux at 4PN for generic orbits, in the center-of-mass (CoM) frame; (ii) the flux at 4.5PN for quasi-circular orbits; and (iii) the $(\ell,\text{m}) = (2,2)$ and $(\ell,\text{m}) = (2,0)$ modes, the latter being purely due to the non-linear memory interactions. The results are all collected in the ancillary file~\cite{AncFile}. App.~\ref{app:BSS} contains the test of the boosted Schwarzchild solution (BSS)~\cite{BDI04zeta}, sucessfully applied to the mass quadrupole moment, and App.~\ref{app:comm_metric} displays the contributions of IR commutators in the 4PN metric.

\section{Non-linear effects in the GW propagation}\label{sec:nonlin}

In this work,\footnote{The conventions employed throughout are as follows: we work with a mostly plus signature; Greek letters denote spacetime indices and Latin letters denote purely spatial indices; we use the multi-index notations, with symmetric and trace-free (STF) mass and current moments $\dM_L \equiv \dM_{i_1i_2\ldots i_\ell}$ and $\dS_L \equiv \dS_{i_1i_2\ldots i_\ell}$; a superscript $(n)$ indicates $n$ time derivatives; hats and angular brackets denote a symmetric trace-free (STF) projection, for instance $\hat{x}_L = x_{\langle L\rangle} = \text{STF}[x_L]$ with $x_L=x_{i_1}\cdots x_{i_\ell}$; the d'Alembertian operator is defined with respect to the flat Minkowski metric $\Box \equiv \eta^{\mu\nu}\partial_{\mu\nu} = \Delta - c^{-2}\partial_t^2$; symmetrization and antisymmetrization are weighted, \emph{e.g.} $A_{(ij)} = (A_{ij} + A_{ji})/2$ and $A_{[ij]} = (A_{ij} - A_{ji})/2$; the usual Levi-Civita symbol is denoted $\epsilon_{ijk}$ with $\epsilon_{123}=1$; finally, as usual, we dubb ``$n$PN'' a quantity of order $\calO(c^{-2n})$.} we compute the GW energy flux of compact binaries on generic orbits and the dominant GW mode $(\ell,\text{m}) = (2,2)$ to the 4PN order, as well as the energy flux on quasi-circular orbits to the 4.5PN order. Let us recall that the general multipole decomposition of the flux reads~\cite{Th80}
\begin{equation}\label{Flux_Thorne}
\mathcal{F} = \sum_{\ell \geqslant 2}\frac{G}{c^{2\ell+1}}\bigg[a_\ell\,\text{U}_L^{(1)}\text{U}_L^{(1)} + \frac{b_\ell}{c^2}\,\text{V}_L^{(1)}\text{V}_L^{(1)}\bigg]\,,
\end{equation}
where $\text{U}_L$ and $\text{V}_L$ (for $\ell\geqslant 2$) respectively denote the mass and current radiative multipole moments (which are STF in their indices), and the numerical coefficients are given by
\begin{equation}\label{eq:aell_bell}
a_\ell = \frac{(\ell+1)(\ell+2)}{(\ell-1)\ell\,\ell!\,(2\ell+1)!!}
\qquad\text{and}\qquad
b_\ell = \frac{4\ell(\ell+2)}{(\ell-1)\,(\ell+1)!\,(2\ell+1)!!}\,.
\end{equation}
The asymptotic waveform, to order $1/R$ in the distance to the source, when expressed in a (Bondi type) radiative coordinate system $(T,X^i)$, with $\bm{X}\equiv R\,\bm{N}$, and written as a perturbation to the ordinary metric $g_{ab}$, reads~\cite{Th80}
\begin{equation}\label{eq:hTT}
h_{ab}^\text{TT} = \frac{4G}{c^2R} \perp_{abij}(\bm{N})\sum_{\ell \geqslant 2}\frac{1}{c^\ell\,\ell!}\bigg[N_{L-2}\,\text{U}_{ijL-2}(u) - \frac{2\ell}{c(\ell+1)}N_{kL-2}\epsilon_{kl(i}\text{V}_{j)lL-2}(u)\bigg] + \calO\left(\frac{1}{R^2}\right)\,,
\end{equation}
where $\perp_{abij}(\bm{N})$ is the usual transverse and traceless (TT) projector and the multipole moments are defined at the retarded time $u \equiv T-R/c$.
The GW amplitude $(\ell,\text{m})$ modes can be read off the radiative moments $\text{U}_L$ and $\text{V}_L$, and for instance the dominant mode $(2,2)$ can be computed directly from the mass quadrupole moment $\text{U}_{ij}$ (in the case of planar, non precessing, orbits; see~\cite{K07,BFIS08,FMBI12}). Therefore, the knowledge of $\text{U}_L$ and $\text{V}_L$ at the relevant PN order will lead to the desired results.

To describe the non-linear effects in the GW propagation from the source to the observer, we connect the radiative moments to the so-called canonical moments $\dM_L$ and $\dS_L$. In turn, the canonical moments are expressed in terms of source moments $\dI_L$ and $\dJ_L$, as well as ``gauge'' moments, $\dW_L$, $\dX_L$, $\dY_L$ and $\dZ_L$. The source and gauge moments describe in our formalism the physics of the PN source. This section is thus devoted to the relations between the radiative moments and the source ones, at the required PN order. We refer to~\cite{B98mult} for a more detailed review and physical discussions about those constructions.

\subsection{Radiative moments versus canonical moments}
\label{subsec:can2rad}

\subsubsection{Radiative moments entering the flux at 4PN order}

In order to derive the energy flux~\eqref{Flux_Thorne} at 4PN order beyond the leading quadrupolar order, the obvious first input is the radiative quadrupole moment itself, $\text{U}_{ij}$, to 4PN order. Recalling that, at leading order, the radiative moments $\text{U}_L$ and $\text{V}_L$ reduce to the $\ell$-th time derivatives (with respect to the retarded time $u$ of the radiative coordinates) of the canonical moments $\dM_L$ and $\dS_L$, we straightforwardly write 
\begin{equation}\label{eq:UijvsMij3D}
\text{U}_{ij} = \dM_{ij}^{(2)} + \text{U}_{ij}^\text{1.5PN}  + \text{U}_{ij}^\text{2.5PN} + \text{U}_{ij}^\text{3PN} + \text{U}_{ij}^\text{3.5PN} + \text{U}_{ij}^\text{4PN} + \mathcal{O}\left(\frac{1}{c^9}\right)\,,
\end{equation}
with small PN corrections up to 4PN, as indicated. The leading correction at 1.5PN is due to the quadratic interaction between the static ADM mass $\dM$ and the mass quadrupole moment $\dM_{ij}$, denoted ``$\dM\times \dM_{ij}$'' and called the ``tail''. It is well-known and reads explicitly~\cite{BD92}
\begin{align}\label{Uij15PN}
\text{U}_{ij}^\text{1.5PN} = \frac{2 G \dM}{c^3} \int_0^{+\infty} \dd\tau\, \dM_{ij}^{(4)}(u-\tau)\bigg[\ln \left(\frac{c \tau}{2 b_0}\right)+ \frac{11}{12}\biggl]\,.
\end{align}
Here the length scale $b_0$ is an arbitrary constant linked with the choice of the origin of time of the asymptotic radiative coordinates, with respect to the harmonic coordinates covering the source's near zone. At the next 2.5PN order, there arises the non-local interaction involving two quadrupole moments $\dM_{ij}\times \dM_{kl}$, called the (displacement) ``memory'', together with associated instantaneous quadrupole-quadrupole terms~\cite{B98quad}, and an instantaneous interaction $\dM_{ij} \times \dS_k$ sometimes dubbed ``failed tail''~\cite{Foffa:2019eeb}:
\begin{equation}\label{Uij25PN}
\text{U}_{ij}^\text{2.5PN} = \frac{G}{c^5} \left\lbrace - \frac{2}{7} \int_0^{+\infty} \! \dd\tau\!\left[\dM^{(3)}_{a\langle i}\dM^{(3)}_{j\rangle a}\right]\!(u-\tau)+ \frac{1}{7}\,\dM^{(5)}_{a\langle i}\dM^{}_{j\rangle a} - \frac{5}{7} \,\dM^{(4)}_{a\langle i}\dM^{(1)}_{j\rangle a} -\frac{2}{7}\,\dM^{(3)}_{a\langle i}\dM^{(2)}_{j\rangle a}  + \frac{1}{3}\epsilon^{}_{ab\langle i}\dM^{(4)}_{j\rangle a}\dS^{}_{b} \right\rbrace\,.
\end{equation}
At the 3PN order appears the first cubic interaction, consisting of the interplay between two ADM masses and the quadrupole moment, \emph{i.e.}, $\dM\times \dM\times \dM_{ij}$, rightly called ``tail-of-tail'' and given by~\cite{B98tail,FBI15}
\begin{equation}\label{Uij3PN}
\text{U}_{ij}^\text{3PN} = \frac{2 G^2 \dM^2}{c^6} \!\!\int_{0}^{+\infty} \!\!\!\dd\tau\,\dM_{ij}^{(5)}(u-\tau) \! \left[\ln^2\left(\frac{c \tau}{2b_0}\right) + \frac{11}{6} \ln\left(\frac{c \tau}{2b_0}\right)- \frac{107}{105} \ln\left(\frac{c \tau}{2r_0}\right) + \frac{124627}{44100}\right]\,.
\end{equation}
Note the appearance here of a new constant length scale $r_0$, which should be distinguished from the previously introduced scale $b_0$. The scale $r_0$ is a running scale in the sense of renormalization group theory~\cite{GRoss10,GRR12,almeida2021gravitational}. The coefficient in front of the $\ln r_0$ term in~\eqref{Uij3PN} is exactly the $\beta$-function coefficient associated to the renormalization of the mass quadrupole moment, say $\beta_2 =-\frac{214}{105}$. In the present formalism, this scale originates from the Hadamard regularization scheme. The next-order 3.5PN term  has a structure similar to the 2.5PN one, \emph{i.e.} with some memory type integrals and instantaneous terms. The interactions between moments, still of quadratic nature, are however more complicated:  
\begin{align}\label{Uij35PN}
\text{U}_{ij}^\text{3.5PN}= \frac{G}{c^7} \Bigg\lbrace &
\int_0^{+\infty} \! \dd\tau\!  \left[ - \frac{5}{756} \dM_{ab}^{(4)} \dM_{ijab}^{(4)} - \frac{32}{63} \dS_{a \langle i}^{(3)} \dS_{j\rangle a}^{(3)}\right]\!(u-\tau) \nn\\
&
- \frac{1}{432}	\dM_{ab} \dM_{ijab}^{(7)} + \frac{1}{432} \dM_{ab}^{(1)} \dM_{ijab}^{(6)} - \frac{5}{756} \dM_{ab}^{(2)} \dM_{ijab}^{(5)} + \frac{19}{648}\dM_{ab}^{(3)} \dM_{ijab}^{(4)}+ \frac{1957}{3024} \dM_{ab}^{(4)} \dM_{ijab}^{(3)}  \nn \\
&
+ \frac{1685}{1008}\dM_{ab}^{(5)} \dM_{ijab}^{(2)} + \frac{41}{28} \dM_{ab}^{(6)}\dM_{ijab}^{(1)} + \frac{91}{216} \dM_{ab}^{(7)} \dM_{ijab} - \frac{5}{252} \dM_{ab \langle i} \dM_{j \rangle ab}^{(7)}+ \frac{5}{189} \dM_{ab \langle i}^{(1)} \dM_{j \rangle ab}^{(6)} \nn \\ 
&
 +\frac{5}{126} \dM_{ab \langle i}^{(2)} \dM_{j \rangle ab}^{(5)} + \frac{5}{2268} \dM_{ab \langle i}^{(3)} \dM_{j \rangle ab}^{(4)}
+ \frac{5}{42} \dS_a \dS_{ija}^{(5)} + \frac{80}{63} \dS_{a \langle i} \dS_{j \rangle a}^{(5)} + \frac{16}{63} \dS_{a \langle i}^{(1)} \dS_{j \rangle a}^{(4)} - \frac{64}{63} \dS_{a\langle i}^{(2)} \dS_{j \rangle a}^{(3)} \nn \\ 
%===============
& 
+\epsilon_{ac \langle i} \bigg( \int_0^{+\infty}\!\dd\tau \left[ \frac{5}{42} \dS_{j\rangle cb}^{(4)} \dM_{ab}^{(3)} - \frac{20}{189} \dM_{j \rangle cb}^{(4)} \dS_{ab}^{(3)} \right]\!(u-\tau) \nn\\
&
\qquad \qquad + \frac{1}{168} \dS_{j \rangle	bc}^{(6)} \dM_{ab} + \frac{1}{24} \dS_{j\rangle bc}^{(5)} \dM_{ab}^{(1)}	+ \frac{1}{28} \dS_{j \rangle bc}^{(4)} \dM_{ab}^{(2)}  - \frac{1}{6} \dS_{j \rangle bc}^{(3)} \dM_{ab}^{(3)} + \frac{3}{56} \dS_{j \rangle bc}^{(2)} \dM_{ab}^{(4)} \nn \\ 
&
\qquad \qquad  + \frac{187}{168} \dS_{j \rangle bc}^{(1)} \dM_{ab}^{(5)} + \frac{65}{84} \dS_{j \rangle bc} \dM_{ab}^{(6)} + \frac{1}{189} \dM_{j	\rangle bc}^{(6)} \dS_{ab} - \frac{1}{189} \dM_{j \rangle bc}^{(5)}	\dS_{ab}^{(1)} \nn \\
& 
\qquad\qquad + \frac{10}{189} \dM_{j\rangle bc}^{(4)} \dS_{ab}^{(2)} + \frac{32}{189} \dM_{j \rangle bc}^{(3)} \dS_{ab}^{(3)} + \frac{65}{189} \dM_{j \rangle bc}^{(2)}\dS_{ab}^{(4)} - \frac{5}{189} \dM_{j \rangle bc}^{(1)} \dS_{ab}^{(5)} - \frac{10}{63} \dM_{j \rangle bc}	\dS_{ab}^{(6)} \bigg)\Bigg\rbrace \,.
\end{align}
At the 4PN order appears the ``tail-of-memory'', which is a cubic interaction involving two time-varying quadrupole moments and the mass, \emph{i.e.} $\dM \times \dM_{ij} \times \dM_{kl}$. It also contains a double integration over the two quadrupole moments; see the first line of~\eqref{Uij4PN}. The tail-of-memory comes with many tail-looking interactions with only one integration over the quadrupole moments. Similarly to the fact that the memory came with the failed tail [see~\eqref{Uij25PN}], the tail-of-memory comes along with a cubic interaction involving the constant angular momentum, \emph{i.e.} $\dM \times \dS_i \times \dM_{jk}$, see the last line of~\eqref{Uij4PN}. We have~\cite{TB23}
\begin{align}\label{Uij4PN}
	\text{U}_{ij}^\text{4PN} =\ & \frac{8 G^2 \dM}{7 c^8} \Bigg\{\int_0^{+\infty} \!\dd\rho\,  \dM_{a \langle i}^{(4)}(u-\rho) \int_0^{+\infty} \!\dd \tau\,  \dM_{j \rangle a}^{(4)}(u-\rho-\tau)  \left[ \ln\left(\frac{c\tau}{2 r_0}\right) - \frac{1613}{270} \right]\nn\\
	%%%%%%%%%%%%%%%%%%%%%%%%
	&\qquad\qquad  - \frac{5}{2} \int_0^{+\infty} \!\dd\tau \,  \left[\dM^{(3)}_{a \langle i} \dM^{(4)}_{j \rangle a}\right](u-\tau)\left[\ln\left(\frac{c\tau}{2 r_0}\right)+\frac{3}{2} \ln\left(\frac{c\tau}{2b_0}\right) \right]\nn\\
	&\qquad\qquad  - 3  \int_0^{+\infty} \!\dd\tau\, \left[\dM^{(2)}_{a \langle i}  \dM^{(5)}_{j \rangle a}\right](u-\tau) \left[\ln\left(\frac{c\tau}{2r_0}\right)  +\frac{11}{12} \ln\left(\frac{c\tau}{2b_0}\right)  \right]\nn\\
	&\qquad\qquad  -\frac{5}{2} \int_0^{+\infty} \!\dd\tau \, \left[\dM^{(1)}_{a \langle i}\dM^{(6)}_{j \rangle a}\right](u-\tau) \left[\ln\left(\frac{c\tau}{2r_0}\right)  + \frac{3}{10} \ln\left(\frac{c\tau}{2 b_0}\right) \right]\nn\\
	&\qquad\qquad    - \int_0^{+\infty} \!\dd\tau\,\left[\dM^{}_{a \langle i}\dM^{(7)}_{j \rangle a}\right](u-\tau) \left[\ln\left(\frac{c\tau}{2r_0}\right) - \frac{1}{4}\ln\left(\frac{c\tau}{2 b_0}\right)\right]\nn\\
	%%%%%%%%%%%%%%%%%%%%%%%%%
	&\qquad\qquad   - 2  \dM^{(2)}_{a \langle i} \int_0^{+\infty} \!\dd\tau\,  \dM^{(5)}_{j \rangle a}(u-\tau) \left[ \ln\left(\frac{c\tau}{2r_0}\right)+ \frac{27521}{5040} \right]\nn\\
	&\qquad\qquad  - \frac{5}{2}\,  \dM^{(1)}_{a \langle i} \int_0^{+\infty} \!\dd\tau \, \dM^{(6)}_{j \rangle a}(u-\tau)  \left[\ln\left(\frac{c\tau}{2r_0}\right)+\frac{15511}{3150} \right]\nn\\
	&\qquad\qquad  + \frac{1}{2} \, \dM_{a \langle i} \int_0^{+\infty} \!\dd\tau \,\dM^{(7)}_{j \rangle a}(u-\tau)  \left[ \ln\left(\frac{c\tau}{2r_0}\right)  - \frac{6113}{756}\right]\,  \Bigg\}\nn\\
	& - \frac{2G^2 \dM}{3c^8}\,\dS_{a} \epsilon_{a b \langle i} \int_{0}^{+\infty} \dd\tau\,  \dM_{j\rangle b}^{(6)}(u-\tau) \left[ \ln\left(\frac{c\tau}{2b_0}\right) +2 \ln\left(\frac{c\tau}{2r_0}\right)  + \frac{1223}{1890}  \right]\,.
\end{align}
Note that the ``genuine'' tail-of-memory given by the first term of~\eqref{Uij4PN} can be retrieved by replacing the canonical moment $\dM_{ij}$  in the expression of the memory term of~\eqref{Uij25PN}  by the radiative moment itself, taking into account the dominant tail effect, \emph{i.e.}
\begin{equation}\label{memToM}
	\dM_{ij}^{(2)} \longrightarrow \text{U}_{ij} = \dM_{ij}^{(2)} + \text{U}_{ij}^\text{1.5PN}\,,
\end{equation}
along with a reexpansion at cubic order and an integration by parts. This agrees with the general prediction for memory effects computed using the radiative multipole moments defined at future null infinity, see~\cite{F09,F11}.

Besides the mass quadrupole moment, the expression of the flux~\eqref{Flux_Thorne} at the 4PN order requires the knowledge of moments of higher multipolarity, but evaluated at lower PN orders. At the next multipolar order, we thus need the mass octupole $\text{U}_{ijk}$ and current quadrupole $\text{V}_{ij}$ moments with a 3PN precision:
\begin{subequations}
\begin{align}	
\text{U}_{ijk} &= \dM_{ijk}^{(3)} + \text{U}_{ijk}^\text{1.5PN}  + \text{U}_{ijk}^\text{2.5PN} + \text{U}_{ijk}^\text{3PN} + \mathcal{O}\left(\frac{1}{c^7}\right)\,,\\
\text{V}_{ij} &= \dS_{ij}^{(2)} + \text{V}_{ij}^\text{1.5PN}  + \text{V}_{ij}^\text{2.5PN} + \text{V}_{ij}^\text{3PN} + \mathcal{O}\left(\frac{1}{c^7}\right)\,.
\end{align}
\end{subequations}
Reporting the results of~\cite{FBI15}, the above contributions read
\begin{subequations}
	\begin{align}
		\text{U}_{ijk}^\text{1.5PN} = 
		&
		\frac{2 G \dM}{c^3} \int_0^{+\infty} \dd\tau\, \dM_{ijk}^{(5)}(u-\tau)\bigg[\ln \left(\frac{c \tau}{2 b_0}\right)+ \frac{97}{60}\biggl]\,,\\
		%==================================
		\text{U}_{ijk}^\text{2.5PN} =\ 
		&
		\frac{G}{c^5} \Bigg\lbrace
		\int_0^{+\infty} \! \dd\tau\left[-\frac{1}{3}\dM^{(3)}_{a\langle i}\dM^{(4)}_{jk\rangle a}- \frac{4}{5}\epsilon_{ab\langle i}\dM^{(3)}_{j\underline{a}}\dS^{(3)}_{k\rangle b}\right]\!(u-\tau)\nn\\
		&\qquad
		+ \frac{1}{4}\dM^{}_{a\langle i}\dM^{(6)}_{jk\rangle a}
		+ \frac{1}{4}\dM^{(1)}_{a\langle i}\dM^{(5)}_{jk\rangle a}
		+ \frac{1}{4}\dM^{(2)}_{a\langle i}\dM^{(4)}_{jk\rangle a}
		- \frac{4}{3}\dM^{(3)}_{a\langle i}\dM^{(3)}_{jk\rangle a}\nn\\
		&\qquad
		- \frac{9}{4}\dM^{(4)}_{a\langle i}\dM^{(2)}_{jk\rangle a}
		- \frac{3}{4}\dM^{(5)}_{a\langle i}\dM^{(1)}_{jk\rangle a}
		+ \frac{1}{12}\dM^{(6)}_{a\langle i}\dM^{}_{jk\rangle a}
		+ \frac{12}{5}\dS^{}_{\langle i}\dS^{(4)}_{jk\rangle}\nn\\
		&\qquad
		+ \epsilon_{ab\langle i}\bigg[
		\frac{9}{5}\dM^{}_{j\underline{a}}\dS^{(5)}_{k\rangle b}
		+ \frac{27}{5}\dM^{(1)}_{j\underline{a}}\dS^{(4)}_{k\rangle b}
		+ \frac{8}{5}\dM^{(2)}_{j\underline{a}}\dS^{(3)}_{k\rangle b}
		+ \frac{12}{5}\dM^{(3)}_{j\underline{a}}\dS^{(2)}_{k\rangle b}\nn\\
		&\qquad\qquad\qquad
		+ \frac{3}{5}\dM^{(4)}_{j\underline{a}}\dS^{(1)}_{k\rangle b}
		+ \frac{1}{5}\dM^{(5)}_{j\underline{a}}\dS^{}_{k\rangle b}
		+ \frac{9}{20}\dM^{(5)}_{jk\rangle a}\dS^{}_b
		\bigg]
		\Bigg\rbrace\,,\\
		%==================================
		\text{U}_{ijk}^\text{3PN} =\
		&
		\frac{2 G^2 \dM^2}{c^6} \!\!\int_{0}^{+\infty} \!\!\!\dd\tau\,\dM_{ijk}^{(6)}(u-\tau) \! \left[\ln^2\left(\frac{c \tau}{2b_0}\right) + \frac{97}{30} \ln\left(\frac{c \tau}{2b_0}\right)- \frac{13}{21} \ln\left(\frac{c \tau}{2r_0}\right) + \frac{13283}{8820}\right]\,,
		\end{align}
where the underlined indices within angled brackets are excluded from the STF operation, and
\begin{align}
		\text{V}_{ij}^\text{1.5PN} =\ 
		&
		\frac{2 G \dM}{c^3} \int_0^{+\infty} \dd\tau\, \dS_{ij}^{(4)}(u-\tau)\bigg[\ln \left(\frac{c \tau}{2 b_0}\right)+ \frac{7}{6}\biggl]\,,\\
		%==================================
		\text{V}_{ij}^\text{2.5PN} =\ 
		&
		\frac{G}{c^5} \Bigg\lbrace
		- \frac{3}{7}\dM^{}_{a\langle i}\dS^{(5)}_{j\rangle a}
		- \frac{3}{7}\dM^{(1)}_{a\langle i}\dS^{(4)}_{j\rangle a}
		+ \frac{8}{7}\dM^{(2)}_{a\langle i}\dS^{(3)}_{j\rangle a}
		+ \frac{4}{7}\dM^{(3)}_{a\langle i}\dS^{(2)}_{j\rangle a}
		+ \frac{17}{7}\dM^{(4)}_{a\langle i}\dS^{(1)}_{j\rangle a}
		+ \frac{9}{7}\dM^{(5)}_{a\langle i}\dS^{}_{j\rangle a}
		- \frac{1}{28}\dM^{(5)}_{ija}\dS^{}_a\nn\\
		&\qquad
		+ \epsilon_{ab\langle i}\bigg[
		- \frac{15}{56}\dM^{}_{j\rangle ac}\dM^{(6)}_{bc}
		- \frac{113}{112}\dM^{(1)}_{j\rangle ac}\dM^{(5)}_{bc}
		- \frac{353}{336}\dM^{(2)}_{j\rangle ac}\dM^{(4)}_{bc}
		- \frac{3}{14}\dM^{(3)}_{j\rangle ac}\dM^{(3)}_{bc}\nn\\
		&\qquad\qquad\qquad
		+ \frac{5}{168}\dM^{(4)}_{j\rangle ac}\dM^{(2)}_{bc}
		+ \frac{3}{112}\dM^{(5)}_{j\rangle ac}\dM^{(1)}_{bc}
		- \frac{3}{112}\dM^{(6)}_{j\rangle ac}\dM^{}_{bc}
		+ \dS^{(4)}_{j\rangle a}\dS^{}_{b}\bigg]
		\Bigg\rbrace\,,\\
		%==================================
		\text{V}_{ij}^\text{3PN} =\
		&
		\frac{2 G^2 \dM^2}{c^6} \!\!\int_{0}^{+\infty} \!\!\!\dd\tau\,\dS_{ij}^{(5)}(u-\tau) \! \left[\ln^2\left(\frac{c \tau}{2b_0}\right) + \frac{7}{3} \ln\left(\frac{c \tau}{2b_0}\right)- \frac{107}{105} \ln\left(\frac{c \tau}{2r_0}\right) - \frac{13127}{11025}\right]\,.
\end{align}
\end{subequations}

Higher order multipole moments have no cubic contributions at the required PN order. 
At 2PN order, we need the mass hexadecapole $\text{U}_{ijkl}$ as well as the current octupole $\text{V}_{ijk}$, which read
\begin{subequations}
	\begin{align}
\text{U}_{ijkl} &= \dM_{ijkl}^{(4)} + \text{U}_{ijkl}^\text{1.5PN}  + \mathcal{O}\left(\frac{1}{c^5}\right)\,,\\
\text{V}_{ijk} &= \dS_{ijk}^{(3)} + \text{V}_{ijk}^\text{1.5PN}  + \mathcal{O}\left(\frac{1}{c^5}\right)\,,
\end{align}
\end{subequations}
where~\cite{FBI15}
\begin{subequations}
\begin{align}
\text{U}_{ijkl}^\text{1.5PN} =\ 
&
\frac{G}{c^3}\Bigg\lbrace2\dM\,\int_0^{+\infty}\!\! \dd\tau\, \dM_{ijkl}^{(6)}(u-\tau)\bigg[\ln \left(\frac{c \tau}{2 b_0}\right)+ \frac{59}{30}\biggl]
+ \frac{2}{5} \int_0^{+\infty} \!\! \dd\tau\,\left[\dM^{(3)}_{\langle ij}\dM^{(3)}_{kl\rangle}\right]\!(u-\tau)\nn\\
&\qquad
- \frac{21}{5}\dM^{}_{\langle ij}\dM^{(5)}_{kl\rangle}
- \frac{63}{5}\dM^{(1)}_{\langle ij}\dM^{(4)}_{kl\rangle}
- \frac{102}{5}\dM^{(2)}_{\langle ij}\dM^{(3)}_{kl\rangle}\Bigg\rbrace\,,\\
%========================
\text{V}_{ijk}^\text{1.5PN} =\ 
&
\frac{G}{c^3}\Bigg\lbrace2\dM\,\int_0^{+\infty}\!\! \dd\tau\, \dS_{ijk}^{(5)}(u-\tau)\bigg[\ln \left(\frac{c \tau}{2 b_0}\right)+ \frac{5}{3}\biggl]\nn\\
&\qquad
-2 \dM_{\langle ij}^{(4)}\dS_{k\rangle}
- \frac{1}{10}\epsilon_{ab\langle i}\dM^{}_{j\underline{a}}\dM^{(5)}_{k\rangle b}
+ \frac{1}{2}\epsilon_{ab\langle i}\dM^{(1)}_{j\underline{a}}\dM^{(4)}_{k\rangle b}
\Bigg\rbrace\,.
\end{align}
\end{subequations}
Note the appearance of a memory integral at 1.5PN order in $\text{U}_{ijkl}$, in addition to the usual tail one. Finally, we need the moments $\text{U}_{ijklm}$ and $\text{V}_{ijkl}$ at 1PN, as well as $\text{U}_{ijklmn}$ and $\text{V}_{ijklm}$ at Newtonian order, which are trivially given by
\begin{subequations}
\begin{align}
\text{U}_{ijklm} &= \dM_{ijklm}^{(5)} + \mathcal{O}\left(\frac{1}{c^3}\right)\,,
& \text{V}_{ijkl} &= \dS_{ijkl}^{(4)} + \mathcal{O}\left(\frac{1}{c^3}\right)\,,\\
\text{U}_{ijklmn} &= \dM_{ijklmn}^{(6)} + \mathcal{O}\left(\frac{1}{c^3}\right)\,,
& \text{V}_{ijklm} &= \dS_{ijklm}^{(5)} + \mathcal{O}\left(\frac{1}{c^3}\right)\,.
\end{align}
\end{subequations}

\subsubsection{Radiative moments entering the quasi-circular flux at the 4.5PN order}

For quasi-circular orbits, the 4.5PN term in the flux was obtained in~\cite{MBF16}. One could naively think that such a computation would require the complete knowledge of the relations between radiative and canonical moments, as presented above, but pushed one half PN order further. This was actually not the case, since for circular orbits, only a limited control of the relation between the radiative and canonical mass quadrupole moments was necessary. This is discussed in~\cite{MBF16}, and we only remind the key points. First, it is well known that the contributions of instantaneous interactions entering the flux at half-integer PN order (\emph{e.g.} 4.5PN order) vanish for quasi-circular orbits. So only non-local contributions such as tails can potentially contribute. Second, the quadratic non-local memory interaction that enters the radiative moments [see Eqs.~\eqref{Uij25PN} and~\eqref{Uij35PN}] become instantaneous in the flux by virtue of time differentiation, so these will not contribute either. Last, the tails-of-memory and spin-quadrupole tails, which both enter at 4PN, will next contribute at 5PN but not at 4.5PN. This allows to use dimensional arguments to determine the interactions that can contribute to the 4.5PN quasi-circular flux. For the mass quadrupole, only the quartic $\dM^3\times \dM_{ij}$, naturally dubbed ``tails-of-tails-of-tails'', can play a role. It has been computed in~\cite{MBF16}, and reads
\begin{align}
\text{U}_{ij}^\text{4.5PN}{\bigg|}_\text{ToToT} =
\frac{G^3 \dM^3}{c^9} \int_0^{+\infty} \dd\tau \,\dM_{ij}^{(6)}(u-\tau)\Bigg[
&
\frac{4}{3}\ln^3 \left(\frac{c\tau}{2 b_0}\right)+ \frac{11}{3}\ln^2\left(\frac{c\tau}{2 b_0}\right) - \frac{428}{105}\ln\left(\frac{c\tau}{2 b_0}\right)\ln\left(\frac{c\tau}{2 r_0}\right)\nn \\
&
+ \frac{124627}{11025}\ln\left(\frac{c\tau}{2 b_0}\right) - \frac{1177}{315} \ln\left(\frac{c\tau}{2 r_0}\right) + \frac{129268}{33075}+ \frac{428}{315}\pi^2 \Bigg]\,.
\end{align}
The full $\text{U}_{ij}^\text{4.5PN}$ also contains quadratic memory interactions, like those entering at 2.5PN and 3.5PN; see~\eqref{Uij25PN} and~\eqref{Uij35PN}. If, as explained above, those are not needed to compute the flux (and phase) for quasi-circular orbits, they will enter the expression of the $(2,2)$ mode. As they are yet undetermined, they restrict the accuracy we can reach when deriving the $(2,2)$ mode, which is why it is presented in Sec.~\ref{subsec:res_modes} at 4PN and not up to 4.5PN order. In addition, radiation reaction effects at 4.5PN in the mass quadrupole should contribute and are also not under control.

Regarding other moments, the 3.5PN terms of the mass octupole $\text{U}_{ijk}$ and current quadrupole $\text{V}_{ij}$, as well as the 2.5PN terms of the mass hexadecapole $\text{U}_{ijkl}$ and current octupole $\text{V}_{ijk}$, are composed of quadratic memory integrals, which cannot contribute in the 4.5PN quasi-circular flux. The only moments playing a role beyond the mass quadrupole are~\cite{FBI15,MBF16} 
\begin{subequations}
\begin{align}
\text{U}_{ijklm}^\text{1.5PN} =\ 
&
\frac{G}{c^3}\Bigg\lbrace2\dM\,\int_0^{+\infty} \dd\tau\, \dM_{ijklm}^{(7)}(u-\tau)\bigg[\ln \left(\frac{c \tau}{2 b_0}\right)+ \frac{232}{105}\biggl]
+ \frac{20}{21} \int_0^{+\infty} \! \dd\tau\,\left[\dM^{(3)}_{\langle ij}\dM^{(4)}_{klm\rangle}\right]\!(u-\tau)\nn\\
&\qquad
- \frac{15}{7}\dM^{}_{\langle ij}\dM^{(6)}_{klm\rangle}
- \frac{41}{7}\dM^{(1)}_{\langle ij}\dM^{(5)}_{klm\rangle}
- \frac{120}{7}\dM^{(2)}_{\langle ij}\dM^{(4)}_{klm\rangle}
- \frac{710}{21}\dM^{(3)}_{\langle ij}\dM^{(3)}_{klm\rangle}\nn\\
&\qquad
- \frac{265}{7}\dM^{(4)}_{\langle ij}\dM^{(2)}_{klm\rangle}
- \frac{155}{7}\dM^{(5)}_{\langle ij}\dM^{(1)}_{klm\rangle}
- \frac{34}{7}\dM^{(6)}_{\langle ij}\dM^{}_{klm\rangle}
\Bigg\rbrace\,,\\
%========================
\text{V}_{ijkl}^\text{1.5PN} =\ 
&
\frac{G}{c^3}\Bigg\lbrace2\dM\,\int_0^{+\infty}\!\! \dd\tau\, \dS_{ijkl}^{(6)}(u-\tau)\bigg[\ln \left(\frac{c \tau}{2 b_0}\right)+ \frac{119}{60}\biggl]\nn\\
&\qquad
- \frac{11}{6}\dM^{}_{\langle ij}\dS^{(5)}_{kl\rangle}
- \frac{25}{6}\dM^{(1)}_{\langle ij}\dS^{(4)}_{kl\rangle}
- \frac{25}{3}\dM^{(2)}_{\langle ij}\dS^{(3)}_{kl\rangle}
- \frac{35}{3}\dM^{(3)}_{\langle ij}\dS^{(2)}_{kl\rangle}
- \frac{65}{6}\dM^{(4)}_{\langle ij}\dS^{(1)}_{kl\rangle}
- \frac{19}{6}\dM^{(5)}_{\langle ij}\dS^{}_{kl\rangle}
- \frac{11}{12}\dM^{(5)}_{\langle ijk}\dS^{}_{l\rangle}\nn\\
&\qquad
+ \epsilon_{ab\langle i}\bigg[
\frac{1}{12}\dM^{}_{j \underline{a}}\dM^{(6)}_{kl\rangle b}
+ \frac{37}{60}\dM^{(1)}_{j \underline{a}}\dM^{(5)}_{kl\rangle b}
- \frac{5}{12}\dM^{(2)}_{j \underline{a}}\dM^{(4)}_{kl\rangle b}
- \frac{5}{6}\dM^{(3)}_{j \underline{a}}\dM^{(3)}_{kl\rangle b}\nn\\
& \qquad\qquad\qquad
- \frac{11}{12}\dM^{(4)}_{j \underline{a}}\dM^{(2)}_{kl\rangle b}
- \frac{1}{12}\dM^{(5)}_{j \underline{a}}\dM^{(1)}_{kl\rangle b}
+ \frac{3}{60}\dM^{(6)}_{j \underline{a}}\dM^{}_{kl\rangle b}\bigg]
\Bigg\rbrace\,.
\end{align}
\end{subequations}

\subsection{Corrections due to the dimensional regularization of radiative moments}
\label{subsec:renorm}

The results presented in the previous section are correct in the ordinary three-dimensional MPM algorithm. Nevertheless, the treatment of the dynamics of point-masses imposes to use a dimensional regularization scheme, starting at the 3PN order~\cite{DJSdim,BDE04}. For consistency purposes, it is thus required to perform the MPM algorithm in $d$ spatial dimensions, too. As proved in~\cite{GRoss10,MQ4PN_renorm}, this induces divergences in the expression of the radiative multipole moments in terms of the canonical ones. Those divergences are in the form of simple poles, \emph{i.e.} they scale as $1/\varepsilon \equiv 1/(d-3)$. Their cancellation against counter-divergences, arising in the computation of the source multipole moments themselves, has been established in~\cite{MQ4PN_IR,MQ4PN_renorm} and represents a crucial check of the method.

Let us recapitulate how the $d$-dimensional radiative multipole moments required for the 4PN flux differ from their three-dimensional counterparts. As the divergences hits at 3PN, only the mass quadrupole and octupole, as well as the current quadrupole, are affected. In the CoM frame (see~\cite{MQ4PN_renorm} for the complete expressions at 3PN in a generic frame), they are
\begin{subequations}\label{eq:dUL_renorm}
	\begin{align}
		\text{U}_{ij}^{(d)} =\ 
		&
		\text{U}_{ij} - \frac{214}{105}\frac{G^2 \dM^2 }{c^6}\left(\Pi_\varepsilon + \frac{246299}{44940}\right) \dM_{ij}^{(4)}\nn\\
		& + \frac{G^2 \dM}{c^8}\bigg[
		- \frac{4}{7}\left(\Pi_\varepsilon - \frac{1447}{216}\right) \dM^{}_{a\langle i}\dM^{(6)}_{j\rangle a}
		- \frac{32}{7}\left(\Pi_\varepsilon - \frac{17783}{10080}\right) \dM^{(1)}_{a\langle i}\dM^{(5)}_{j\rangle a}\nn\\
		& \qquad\qquad
		- 4\left(\Pi_\varepsilon - \frac{27649}{17640}\right) \dM^{(2)}_{a\langle i}\dM^{(4)}_{j\rangle a}
		+ \frac{1921}{945} \dM^{(3)}_{a\langle i}\dM^{(3)}_{j\rangle a}
		+ \frac{4}{3}\left(\Pi_\varepsilon + \frac{11243}{7560}\right) \dM^{(5)}_{a\langle i}\dS^{}_{j\rangle \vert a}\bigg]\,,
		\label{eq:dUij_renorm}\\
		%==========================
		\text{U}_{ijk}^{(d)} =\ 
		&
		\text{U}_{ijk} - \frac{21}{26}\frac{G^2 \dM^2 }{c^6}\left(\Pi_\varepsilon + \frac{9281}{2730}\right) \dM_{ijk}^{(5)}\,,\label{eq:dUijk_renorm}\\
		%==========================
		\text{V}_{k\vert ji}^{(d)}  =\ 
		&
		\text{V}_{k\vert ji} - \frac{214}{105}\frac{G^2 \dM^2 }{c^6}\left(\Pi_\varepsilon + \frac{4989}{44940}\right) \dS_{k\vert ji}^{(4)}\,,\label{eq:dVij_renorm}
	\end{align}
\end{subequations}
where we employ the notations of~\cite{HFB_courant} for current moments, and where
\begin{equation}
	\Pi_\varepsilon = - \frac{1}{2 \varepsilon} + \ln\left(\frac{r_0 \sqrt{\bar{q}}}{\ell_0}\right)\,,
%	\qquad\text{with}\quad \bar{q} \equiv 4 \pi e^{\gamma_\text{E}}\,.
\end{equation}
with $\bar{q} \equiv 4 \pi \de^{\gamma_\text{E}}$, $\gamma_\text{E}$ being the Euler constant. The constant $\ell_0$ is the length-scale associated with dimensional regularization, such that the $d$-dimensional gravitational coupling constant $G_d = \ell_0^\varepsilon \,G$ relates to the usual Newton's constant $G$, and $r_0$ was introduced in the previous section. As expected, the numerical constants associated with the poles at 3PN are the $\beta$-coefficients of the renormalization group flows for these multipole moments~\cite{GRoss10,GRR12,almeida2021gravitational}.\footnote{See~\cite{Blanchet:2019rjs} for the computation using the renormalization group flow of high-order logarithmic effects in the conservative energy function.}

As explained in Sec.~\ref{sec:source} the poles present in~\eqref{eq:dUL_renorm} will actually be cancelled by poles coming from the $d$-dimensional expression of the source multipole moments. Thus, the correction terms~\eqref{eq:dUL_renorm}, added to the $d$-dimensional source moments, permit to define a notion of finite (``renormalized'') source moment, which is useful in some intermediate steps of our calculation (see
more details in~\cite{MQ4PN_renorm}). 

\subsection{Relation between canonical and source moments}
\label{subsec:source2can}

In the previous sections, the radiative moments have been linked to a set of canonical moments, $\dM_L$ and $\dS_L$. Nevertheless, the matching of the linearized metric, which serves as a basis for the MPM construction, to the PN source is more easily done by using a set of source moments $\dI_L$ and $\dJ_L$, together with four families of gauge moments $\dW_L$, $\dX_L$, $\dY_L$ and $\dZ_L$. We thus need to connect the canonical moments to those source and gauge moments, which is done \emph{via} a coordinate transformation. This coordinate change (\emph{i.e.}, fully non-linear gauge transformation) induces a precise prescription for the relation between the canonical moments and the source moments, which has been worked out at 4PN order in three dimensions in~\cite{MQ4PN_jauge}. However, since we are using full dimensional regularization in this work, we must generalize this computation to $d$ dimensions in order to be consistent. Throughout this section, we will assume that the reader is familiar with the construction and notations of~\cite{MQ4PN_jauge}, and we will only highlight the main differences between the $d$-dimensional and the three-dimensional computations.

Regarding 4.5PN, non-local terms cannot appear at that order in the relation between canonical and source/gauge moments. Indeed, in the three-dimensional case, non-local terms do arise at cubic order, but dimensional analysis shows that there cannot be any cubic or higher-order contribution entering at exactly 4.5PN order. As for quadratic interactions entering at 4.5PN order, an analysis of the structure of the integration formulas proves that they are necessarily instantaneous. Therefore, as discussed previously, they will play no role in the flux for quasi-circular orbits, and we only need the 4PN order in the canonical/source/gauge relations to control the 4.5PN quasi-circular flux.

\subsubsection{Procedure in $d$ dimensions}

The solution of the linearized, $d$ dimensional, vacuum Einstein field equations in harmonic gauge reads
\begin{equation}
h^{\mu\nu}_{\text{gen}\,1} =  h^{\mu\nu}_{\text{can}\,1} + \partial^\mu \varphi^\nu_1 + \partial^\nu \varphi^\mu_1-\partial_\lambda \varphi^\lambda_1 \,\eta^{\mu\nu}\,,
\end{equation}
where the canonical metric is parametrized by three types of source moments as~\cite{HFB_courant}
\begin{subequations}\label{hcan1}
\begin{align}
h^{00}_{\text{can}\,1} 
&= -\frac{4}{c^2}\sum_{\ell\geqslant 0} \frac{(-)^\ell}{\ell!}\partial_L \widetilde{\dI}_L \,, \\
h^{0i}_{\text{can}\,1} 
&= \frac{4}{c^3} \sum_{\ell\geqslant 1} \frac{(-)^\ell}{\ell!}\left[\partial^{}_{L-1} \widetilde{\dI}^{(1)}_{iL-1} + \frac{\ell}{\ell+1}\partial^{}_{L} \widetilde{\dJ}_{i|L}\right] \,, \\ 
h^{ij}_{\text{can}\,1} 
&= -\frac{4}{c^4} \sum_{\ell\geqslant 2} \frac{(-)^\ell}{\ell!}\left[\partial^{}_{L-2} \widetilde{\dI}^{(2)}_{ijL-2} + \frac{2\ell}{\ell+1}\partial^{}_{L-1} \widetilde{\dJ}^{(1)}_{(i|\underline{L-1} j)}+ \frac{\ell-1}{\ell+1}\partial^{}_L \widetilde{\mathrm{K}}_{ij|L}\right] \,, \label{hijcan1}
\end{align}
\end{subequations}
and the gauge vector $\varphi^\mu_1$, by four types of gauge moments,
\begin{subequations}\label{eq:Jauge_vector}
\begin{align}
\varphi^0_1 
&= \frac{4}{c^3}\sum_{\ell\geqslant 0} \frac{(-)^\ell}{\ell!}\partial_L \widetilde{\dW}_L \,, \\
\varphi^i_1 
&= -\frac{4}{c^4}\sum_{\ell\geqslant 0} \frac{(-)^\ell}{\ell!}\partial_{iL}\widetilde{\dX}_L
 -\frac{4}{c^4}\sum_{\ell\geqslant 1} \frac{(-)^\ell}{\ell!}\bigg[\partial_{L-1}\widetilde{\dY}_{iL-1} + \frac{\ell}{\ell+1}\,\partial_L\widetilde{\dZ}_{i|L} \bigg]\,.
\end{align}
\end{subequations}
The conventions, notably for the indices of $\widetilde{\dJ}_{i\vert L}$ and $\widetilde{\dZ}_{i\vert L}$, are still those of~\cite{HFB_courant},\footnote{
There has been some confusion~\cite{HFB_courant, MQ4PN_jauge, TLB22} concerning these conventions, which we clarify here. In the~HFB~convention~\cite{HFB_courant}, the ordering of the multi-index in the multipolar moments is defined in an unusual order by $\dJ_{i|L} \equiv \dJ_{i|i_\ell ... i_1}$ and  $\mathrm{K}_{ij|L} \equiv \mathrm{K}_{ij|i_\ell ... i_1}$. In the BFL convention \cite{MQ4PN_jauge} however, the multi-index is always defined in the usual order, namely $\dJ_{i|L} \equiv \dJ_{i|i_1 ... i_\ell}$ and  $\mathrm{K}_{ij|L} \equiv \mathrm{K}_{ij|i_1 ... i_\ell}$. Note that the mass-type moment is entirely symmetric, so the ordering of its indices does not matter, namely $\dI_{L} = \dI_{i_1 ...i_\ell} = \dI_{i_\ell ... i_1}$. In both conventions, the Young tableaux of the moments read (with the notation of~\cite{MQ4PN_jauge}):
\begin{align*}
	\dI_L =\ytableausetup{boxsize=1.5em,textmode}
	\ytableaushort{{\scriptsize $i_\ell$} {...} {\scriptsize $i_1$}}\,,\qquad \qquad\qquad
	\dJ_{i\vert L}=
	\ytableaushort{{\scriptsize $i_\ell$} {\scriptsize
			$\,i_{\scalebox{0.6}{\text{$\ell\! -\! 1$}}}$} {...} {\scriptsize 
			$i_1$}, {\scriptsize $i$}}~\,, \qquad\qquad\qquad
	\mathrm{K}_{ij\vert L} = \ytableaushort{{\scriptsize $i_\ell$} {\scriptsize
			$\,i_{\scalebox{0.6}{\text{$\ell\!- \! 1$}}}$} {\scriptsize
			$i_{\scalebox{0.6}{\text{$\ell\!- \! 2$}}}$} {...} {\scriptsize $i_1$},
		{\scriptsize $j$} 
		{\scriptsize $i$}}~\,.
\end{align*}
The current moment can be expressed using a Levi-Civita symbol in the $d \rightarrow 3$ limit. This relation is in both conventions: $\dJ_{i|L} = \epsilon_{i i_\ell a} \dJ_{a L-1}$, where $\dJ_L$ is symmetric in its indices. Note also that the conventions for $\dZ_{i|L}$ are the same as for $\dJ_{i|L}$. In practice, the only difference between the two conventions is that the object ${\dJ}_{(i|\underline{L-1} j)}$ of~\eqref{hijcan1}, valid in the HFB convention (where the underlined indices are excluded from the symmetrization), would read instead ${\dJ}_{(i|j) L-1}$ in the BFL convention; see for example (2.5c) of~\cite{MQ4PN_jauge}.
} 
and we have shortened
\begin{subequations}\label{eq:Jauge_Ftilde_def}
	\begin{align}
	&\widetilde{F}(t,r) = \frac{\widetilde{k}}{r^{d-2}} \int_1^{+\infty} \dd z\, \gamma_{\frac{1-d}{2}}(z) \,F(t-zr/c)\,,\\
	\text{with}\quad &\gamma_s(z) \equiv \frac{2\sqrt{\pi}}{\Gamma(1+s)\Gamma(-\frac{1}{2}-s)}\,\big(z^2-1\bigr)^s\quad\text{and}\quad\tilde{k} \equiv \frac{\Gamma\left(\frac{d-2}{2}\right)}{\pi^\frac{d-2}{2}}\,.
	\end{align}
\end{subequations}
This shorthand satisfies $\lim_{d\to3}\widetilde{F}(t,r) = F(t-r/c)/r$. Note the appearance of a new type of moment, $\mathrm{K}_{ij\vert L}$, which is a pure artifact of working in $d \neq 3$ dimensions, and will not enter our results since it has zero independent components in three dimensions, see App.~A of~\cite{HFB_courant}, where the number of independent components is given by~(A6b).

Starting from this $d$-dimensional metric, we can follow the lines of Sec.~III.A in~\cite{MQ4PN_jauge}, which are independent of the dimension. Notably, at any PM order $n$, the expressions of the non-linear interaction terms $\Omega_n^{\mu\nu}$ and $\Delta^\mu_n$ as functions of $h^{\mu\nu}_{\text{can}\,n}$ and $\varphi_n^\mu$ are identical to the three-dimensional case. The main, but minor, difference concerns the two key quantities $X_n^{\mu\nu}$ and $Y_n^{\mu\nu}$. If their formal definition, Eqs.~(3.20) of~\cite{MQ4PN_jauge}, does not depend of the dimension, their explicit expressions in terms of $\Omega_n^{\mu\nu}$ and $\Delta_n^\mu$ are slightly changed, and they read in $d$ dimensions
\begin{subequations}\label{eq:defXnYn}
\begin{align}
X_n^{\mu\nu}
&= \FP \Box_\text{ret}^{-1}\left[B \left(\frac{r}{r_0}\right)^B\left(- \frac{B+d-2}{r^2} \,\Omega_n^{\mu\nu} - \frac{2}{r}\partial_r \Omega_n^{\mu\nu}\right)\right]\,,\\
Y_n^{\mu\nu} 
&= \FP \Box_\text{ret}^{-1}\left[B \left(\frac{r}{r_0}\right)^B \frac{n_k}{r}\left( - 2 \delta^{k( \mu} \Delta_n^{\nu)} + \eta^{\mu\nu} \Delta_n^k \right)\right]\,.
\end{align}
\end{subequations}
These quantities are endowed with a Hadamard finite part (FP) regularization when $B\to 0$, here defined on top of the dimensional regularization. Note the explicit prefactor $B$ in the source of the retarded d'Alembertian operators in~\eqref{eq:defXnYn}. The quantities $X_n^{\mu\nu}$ and $Y_n^{\mu\nu}$ will be non-zero only when the retarded integrals develop a pole $\propto B^{-1}$. The remaining of the procedure described in Sec.~III of~\cite{MQ4PN_jauge}, and notably the extraction of the correction to the canonical moments, is left unchanged. All the difference will reside in the detailed integration formulas to compute~\eqref{eq:defXnYn}.

\subsubsection{Integration techniques}

Let us recall that the solution at an order $n\geqslant 2$ is composed of some corrections to the canonical moments, extracted from $X_n^{\mu\nu}$ and $Y_n^{\mu\nu}$ given by~\eqref{eq:defXnYn}, as well as a gauge vector, denoted $\varphi_n^\mu$. Thanks to the prefactor $B$ in~\eqref{eq:defXnYn}, and similarly to the three-dimensional case, the computation of $X_n^{\mu\nu}$ and $Y_n^{\mu\nu}$ only requires the near-zone ($r\to 0$) behavior of the functions $\Omega_n^{\mu\nu}$ and $\Delta_n^\mu$, as the FP regularization is introduced to cope with the singular behavior of multipole expansions when $r\to 0$. Using the expansion formulas displayed in App.~A of~\cite{BBBFMc}, the computation boils down to the evaluation of integrals of the form
\begin{equation}\label{eq:gauge_I_def}
{\cal J}^b_{p,q} \equiv \FP \Box_{\text{ret}}^{-1}\left[\left(\frac{r}{r_0}\right)^B B^b  \frac{\hat{n}_L}{r^{p+q \varepsilon}}\, H\left(t\right) \right]\,,
\end{equation}
where $b=1,2$ and $H$ can bear simple poles $\propto 1/\varepsilon$ and can indifferently be a local or non-local function of time. In this section, we set $r_0 = \ell_0 = 1$ for simplicity. The integrals~\eqref{eq:gauge_I_def} are nothing but the $d$-dimensional generalization of~(4.13) of~\cite{MQ4PN_jauge}, with $a=0$, since the case which includes a $\ln r$ is irrelevant for the present generalization.

Techniques to solve wave-like equations in $d$ dimensions, with more general radial dependence than~\eqref{eq:gauge_I_def} but same multipolarity $\ell$, have been developed in~\cite{MQ4PN_renorm}, see Sec.~II therein. Importantly, no particular assumption about the presence of a prefactor $B^b$ has been made in this work, as clear from its Eq.~(2.6). We can therefore follow the lines of~\cite{MQ4PN_renorm} to compute the integral~\eqref{eq:gauge_I_def}. The solution $h$ is the sum of two contributions, $h^{<}$ and $h^{>}$, involving integrals over the respective domains $\mathcal{D}=\{|\mathbf{x}'|<r\}$ and $\mathbb{R}^3-\mathcal{D}$, as displayed explicitly in Eqs.~(2.8) and~(2.10) of~\cite{MQ4PN_renorm}. When a prefactor $B^b$ is present, with $b=1,2$, the finite part of $h\equiv \mathcal{J}^b_{p,q}$ vanishes unless the integration produces a pole, which may only occur when the integral diverges for $B=0$. As the system is taken to be stationary in the past in the MPM algorithm, the integral yielding $h^{>}$ converges for $B=0$ so that only $h^{<}$ can contribute to the finite part in Eq.~\eqref{eq:gauge_I_def}. This finite part is evaluated explicitly and truncated at first order in $\varepsilon$, in order to cope with the case where $H(t)$ contains a pole, namely $H=\frac{1}{\varepsilon}H_{-1}+H_0+\calO(\varepsilon)$. After resummation,
we obtain a retarded homogeneous solution of the wave equation~\cite{MQ4PN_renorm}
\begin{equation}
{\cal J}^b_{p,q} =  \hat{\partial}_L\widetilde{G}^b_{p,q}(t,r) \equiv \hat{\partial}_L\,\left[\frac{\tilde{k}}{r^{d-2}} \,\int_1^{+\infty} \dd z \,\gamma_{\frac{1-d}{2}}(z) \,G^b_{p,q}(t-z r/c)\right]\,.
\end{equation}
Explicitly, we find that $G^b_{p,q}$ is non-vanishing only if $b = 1$, $q=1$ and $p-\ell-3 = 2j$, where $j \in \mathbb{N}$. Under those very restrictive conditions, we have
\begin{equation}\label{G(t)}
	G^b_{p,q}(t) = \frac{(-)^{\ell+1}}{(2j)!!\,(2j+2\ell+1)!!}\left[\left(1 + \varepsilon \ln\sqrt{\bar{q}} - \varepsilon \sum_{k=0}^{j+\ell}\frac{1}{2k+1}\right)H^{(2j)}(t) + \calO(\varepsilon)\right]\,,
\end{equation}
where we recall that $H$ can bear a pole $1/\varepsilon$, and our notation $\bar{q}=4\pi\de^{\gamma_\text{E}}$ (with $\gamma_\text{E}$ the Euler constant). With the previous formula at hand, we are ready to compute the quantities $X_n^{\mu\nu}$ and $Y_n^{\mu\nu}$ and, therefore, obtain the correction to the canonical moments to the $n$-th PM order. Since the result~\eqref{G(t)} is zero unless $q=1$, we are able to eliminate many terms from the source for this computation solely based on the value of $q$.

As for the gauge vector $\varphi_n^\mu$, it will be needed to compute the source term for the next PM order $n+1$, following the algorithmic procedure of~\cite{MQ4PN_jauge}. It is decomposed as $\varphi_n^\mu = \phi_n^\mu + \psi_n^\mu$, where $\psi_n^\mu$ is also extracted from  $X_n^{\mu\nu}$ and $Y_n^{\mu\nu}$, and $\phi_n^\mu$ is obtained by the direct integration
\begin{equation}
\phi_n^\mu \equiv \FP \Box_\text{ret}^{-1}\left[\left(\frac{r}{r_0}\right)^B \Delta_n^\mu\right] \,.
\end{equation}
In $d$ dimensions, it is in general not possible to find a convenient, explicit expression for $\phi_n^\mu$ (see, however, Sec.~II in~\cite{MQ4PN_renorm}). Fortunately, we do not need here the full solution valid at any field point, but only the solution in the form of a near-zone expansion when $r\to 0$. Indeed, following the procedure of~\cite{MQ4PN_jauge}, it is the near-zone expansion of $\phi_n^\mu$, denoted $\bar{\phi}_n^\mu$, that needs to be inserted into the expressions of $\Omega_{n+1}^{\mu\nu}$ and $\Delta^\mu_{n+1}$ sourcing the next order quantities $X_{n+1}^{\mu\nu}$ and $Y_{n+1}^{\mu\nu}$. The quantity $\bar{\phi}_n^\mu$ can be obtained directly from the near-zone expansion of the source term, \emph{i.e.} $\bar{\Delta}_n^\mu$, plus a crucial homogeneous solution, itself in the form of a near-zone expansion, namely:
\begin{equation}
	 \bar{\phi}_{n}^\mu=\bar{\phi}_{\text{part}\,n}^\mu+\bar{\phi}_{\text{hom}\,n}^\mu\,.
\end{equation}
The particular solution is obtained by a formal iteration of inverse Laplace operators in $d$-dimensions, using the ``Matthieu'' formula, Eq.~(B.26c) in~\cite{BDE04}:
\begin{equation}
	\bar{\phi}_{\text{part}\,n}^\mu  = \FP \,\sum_{k=0}^{+\infty}\Delta^{-1-k}\left(\frac{\partial}{c\,\partial t}\right)^{2k}\left[\left(\frac{r}{r_0}\right)^B \bar{\Delta}_n^\mu\right] \,.
\end{equation}
The homogeneous solution $\bar{\phi}_{\text{hom}\,n}^\mu$ is given by Eq.~(3.20) of~\cite{BBBFMc}, which in this case yields (with $c=1$)
\begin{equation} 
	\bar{\phi}_{\text{hom}\,n}^\mu  = -\sum_{\ell\geqslant 0} \sum_{j=0}^{+\infty} \frac{ \Delta^{-j}\hat{x}_L}{d+2\ell-2} \FP \int_1^{+\infty} \dd z\, \gamma_{\frac{1-d}{2}-\ell}(z) \int_0^{+\infty} \dd r' r'^{-\ell+1+B} \,\hat{n}'_L  \,\bar\Delta_{n,L}^{\mu\, (2j)}(t-z r', r') \,,
\end{equation}
where we have decomposed the source in its multipolar components as
\begin{equation} \bar\Delta_n^\mu(t,\mathbf{x}) = \sum_{\ell\geqslant 0} \hat{n}_L  \,\bar\Delta_{n,L}^\mu(t,r) \,,\end{equation}
and shortened the iterated inverse Laplacians as~\cite{PB02,BFN05,BBBFMc}
\begin{equation}\label{notationLaplace}
\Delta^{-j} \hat{x}_L  = \frac{\Gamma\left(\frac{d}{2}+\ell\right)}{\Gamma\left(\frac{d}{2}+\ell+j\right)} \frac{r^{2j}\hat{x}_L}{2^{2j}\,j!}\,.
\end{equation}
In the particular case where the source term has a structure given by 
\begin{equation} \label{IntegrableStructureSource}
\bar\Delta_{n}^\mu(t,r)  = \sum_{\ell, k, p, q} \frac{\hat{n}_L}{r^{p+q \varepsilon}} \int_1^{+\infty} \dd y\, y^k \gamma_{-1-\frac{\varepsilon}{2}}(y) \,F_{\ell,k,p,q}(t-y r) \,,
\end{equation}
where the $F_{\ell,k,p,q}(t)$'s can be arbitrary functions of time, explicit formulas are known for performing the integrations: see (3.24) and (D1) of~\cite{BBBFMc} when $q=2$, and the discussion in Sec. IV of~\cite{MQ4PN_renorm} for other values of $q$. However, we will encounter source terms that exhibit more complicated structures, and for which no analogous formula is available. This is notably the case for sources involving two non-static multipolar moments, such as $\Delta_{\dW \times \dI_{ij}}^\mu$. Fortunately, we have shown that those more complicated homogeneous solutions do not contribute at 4PN order, see hereafter.

\subsubsection{The 4PN relation between canonical and source moments}

At the 4PN order, only quadratic and cubic interactions are allowed by dimensional analysis (and we recall that the 4.5PN sector vanishes in the quasi-circular flux). Naturally, the moments allowed to enter those interactions do not depend on the space dimension. Applying the method described above, it was found that the quadratic order is identical in $d$ and three dimensions, \emph{i.e.} we recover the same ``odd'' corrections at 2.5PN and 3.5PN orders in the relation between canonical and source/gauge moments. More interestingly, the results for the cubic interactions arising at 4PN order greatly differ.

Let us recall that three interactions enter the 4PN mass quadrupole moment at cubic order, namely $\dM\times \dM\times \dW_{ij}$ and $\dM\times \dM\times \dY_{ij}$ (where only one moment is non-static), and most importantly $\dM\times \dW\times \dI_{ij}$ (where the two moments $\dW$ and $\dI_{ij}$ are dynamical). For the first two interactions, it turns out that the source term entering the integral~\eqref{eq:gauge_I_def} bears $q \geqslant 2$. Indeed, it is composed of two interactions: $h_{\dM \times \dM} \times \varphi_{\mathrm{K}_{ij}}$ and $h_{\dM} \times \varphi_{\dM \times \mathrm{K}_{ij}}$, where $\mathrm{K}_{ij}$ stands for either $\dW_{ij}$ or $\dY_{ij}$. As both $h_{\dM \times \dM}$ and $\varphi_{\dM \times \mathrm{K}_{ij}}$ bear $q=2$, the cubic sources will have $q \geqslant 2$, and thus cannot satisfy the $q=1$ necessary condition to have a non-vanishing function $G(t)$ in Eq.~\eqref{G(t)}. Therefore, the $\dM\times \dM\times \dW_{ij}$ and $\dM\times \dM\times \dY_{ij}$ interactions do not contribute in $d$ dimensions, as we have explicitly checked, using the method presented in the previous sections. This vanishing of the  $\dM\times \dM\times \dW_{ij}$ and $\dM\times \dM\times \dY_{ij}$ interactions in $d$ dimensions contrasts with their explicit contributions in 3 dimensions; see Eq.~(1.1) in~\cite{MQ4PN_jauge}.

As for the last cubic interaction, $\dM\times \dW\times \dI_{ij}$, it turns out that the source term \emph{does} have a $q=1$ component, arising from the interaction $h_{\dI_{ij}} \times \varphi_{\dM\times \dW}$. More precisely, the term with $q=1$ comes exclusively from the interaction between: (i) the homogeneous solution $\bar{\phi}_{\text{hom}\,2}$ at quadratic order arising from the interaction $\dM\times \dW$, \emph{i.e.} $\bar{\phi}_{\text{hom} \,\dM\times \dW}$, which bears $q=0$; and (ii) the $q=1$ piece of the linear metric $h_{\dI_{ij}}$; see~\eqref{hcan1}. 
Note that, as it involves a quadratic interaction with only one dynamical moment, $\varphi_{\dM\times \dW}$ can be calculated using~\eqref{IntegrableStructureSource}.
By contrast, the homogeneous solution for the interaction $\dW \times \dI_{ij}$, \emph{i.e.}  $\bar{\phi}_{\text{hom}\,\dW \times \dI_{ij}}$, does not fulfill the $p-\ell-3 \in 2\mathbb{N}$ condition, and as such, does not contribute to the result. This is fortunate, since it does not follow the structure~\eqref{IntegrableStructureSource}, and we are unable to compute this term explicitly, at least with currently-known formulas.
Another interesting observation is that the explicit computation of the $q=1$ terms (the only ones that contribute to the final result) leads to the appearance of a pole. The relation between canonical and source/gauge moments at 4PN is thus a pure $\dM\times \dW\times \dI_{ij}$ interaction, which reads
\begin{equation}\label{eq:gauge_dMij_ddim}
\delta \dM_{ij} = \frac{8\,G^2\dM}{c^8}\int_0^{+\infty}\!\dd\tau \bigg[\frac{1}{\varepsilon}-2\ln\left(\frac{c\tau \sqrt{\bar{q}}}{2}\right) - 1\bigg]\bigg(\dI_{ij}^{(1)}(t)\,\dW^{(3)}(t-\tau)-\dI_{ij}(t)\,\dW^{(4)}(t-\tau)\bigg)+ \mathcal{O}\left(\frac{1}{c^{10}}\right)\,.
\end{equation}
This result is consistent with the corresponding one in three dimensions, where the interaction $\dM\times \dW\times \dI_{ij}$ gives rise to an ordinary tail integral at 4PN order: the coefficient of the pole in~\eqref{eq:gauge_dMij_ddim} matches the logarithmic structure of Eq.~(1.1) in~\cite{MQ4PN_jauge}. Moreover, we have checked the result~\eqref{eq:gauge_dMij_ddim}, as well as the vanishing of the interactions $\dM\times \dM\times \dW_{ij}$ and $\dM\times \dM\times \dY_{ij}$, by an independent method, using the techniques exposed in~\cite{MQ4PN_renorm}, to compute the \emph{difference} between the $d$-dimensional and Hadamard regularization schemes in the MPM algorithm, and adding it to the three-dimensional result of~\cite{MQ4PN_jauge}.

The new pole in~\eqref{eq:gauge_dMij_ddim} arising from dimensional regularization is actually cancelled by another pole. This stems from an extra contribution to be added to the end result of~\cite{MQ4PN_IR}, which we will now discuss. Indeed, when computing the equations of motion at 4PN order, the tail effect (entering as a radiation mode in the conservative dynamics described by the Fokker action) has been implemented using the canonical quadrupole moment~$\dM_{ij}$~\cite{BBBFMc}, and is therefore valid in the associated ``canonical'' coordinate system. By contrast, the source quadrupole moment $\dI_{ij}$, as given in~\cite{MQ4PN_IR}, is expressed in a so-called ``generic'' coordinate system, following the terminology of~\cite{MQ4PN_jauge}. Note that the computation of the source quadrupole moment $\dI_{ij}$ at 4PN does not use the 4PN equations of motion. These two results, taken independently, are therefore perfectly correct in their respective coordinate systems. However, in this work, we need to take time derivatives of the source quadrupole moment, which crucially requires the 4PN equations of motion. This operation can only be performed if the source quadrupole moment and the equations of motion are expressed in the same coordinate system. We thus compute the expression of the source quadrupole moment in the canonical coordinate system rather than in the generic one. This is done by performing a simple coordinate change $x^\mu_\text{gen} = x^\mu_\text{can} + \zeta^\mu$, which transforms the gravitational field according to Eqs.~(3.1)--(3.3) in~\cite{MQ4PN_jauge}. A simple dimensional analysis shows that only the $\dM\times \dW$ interaction can enter this coordinate change at 4PN order. Using the methods exposed previously to calculate the contribution of a given interaction to the gauge vector, we find that
\begin{subequations}
\begin{align}
\zeta^0 &\equiv \varphi^0_{\text{hom}\,\dM\times \dW} = \frac{8\,G^2\dM}{c^{7}} \sum_{j\geqslant 0} \frac{\Gamma\left(\frac{d}{2}\right)}{\Gamma\left(\frac{d}{2}+j\right)}\frac{r^{2j}}{2^{2j}\,j!\,c^{2j}}\int_0^{+\infty}\!\dd\tau \bigg[\frac{1}{\varepsilon}-2\ln\left(\frac{c\tau\,\sqrt{\bar{q}}}{2}\right) - 1\bigg]\dW^{(3+2j)}(t-\tau)\,,\\
\zeta^i &\equiv \varphi^i_{\text{hom}\,\dM\times \dW} = 0\,,
\end{align}
\end{subequations}
where only $j=0$ is relevant at 4PN. Implementing the procedure exposed in Sec. II of~\cite{MQ4PN_IR} with the ``pure gauge'' metric $h^{\mu\nu}=- \partial^\mu\zeta^\nu-\partial^\nu\zeta^\mu+\eta^{\mu\nu}\partial_\lambda\zeta^\lambda$, we arrive at
\begin{equation}\label{eq:gauge_dMij_ddim2}
	\delta_\zeta\dI_{ij} = - \frac{8\,G^2\dM}{c^8}\int_0^{+\infty}\!\dd\tau \bigg[\frac{1}{\varepsilon}-2\ln\left(\frac{c\tau \sqrt{\bar{q}}}{2}\right) - 1\bigg]\bigg(\dI_{ij}^{(1)}(t)\,\dW^{(3)}(t-\tau)-\dI_{ij}(t)\,\dW^{(4)}(t-\tau)\bigg)+ \mathcal{O}\left(\frac{1}{c^{10}}\right)\,.
\end{equation}
This effect is formally a correction to the \emph{source} mass quadrupole moment $\dI_{ij}$. 
Nevertheless, it is more handy to identify it as a correction in the relation between the canonical and source/gauge moments. At the 4PN order, no obstacle interferes with such identification. Thus, we will consider it as an ``indirect'' contribution in this relation, namely $\delta_\text{indirect}\dM_{ij}\equiv \delta_\zeta\dI_{ij}$, to be added to the ``direct'' one given by~\eqref{eq:gauge_dMij_ddim}. We then find that, at least at 4PN order, the direct and indirect contributions exactly cancel:
\begin{equation}\label{directindirect}
	\delta_\text{direct}\dM_{ij} + \delta_\text{indirect}\dM_{ij} = 0\,.
\end{equation}
Therefore, we have proved that all cubic interactions in $d$ dimensions are vanishing at 4PN order, in clear contrast with the three-dimensional result, Eq.~(1.1) of~\cite{MQ4PN_jauge}. Finally, as there is no pole left in the final relation between the source and canonical quadrupole, we can reduce it directly to three dimensions. This relation, using full dimensional regularization, reads
\begin{align}\label{eq:resMij}
\dM_{ij} &= 
	\dI_{ij}
	+ \frac{4G}{c^5}\bigg[\dW^{(2)}\dI^{}_{ij}-\dW^{(1)}\dI^{(1)}_{ij}\bigg]\nn\\
	&\
	+ \frac{4G}{c^7}\Bigg\{
	\frac{4}{7}\dW^{(1)}_{a\langle i}\dI^{(3)}_{j\rangle a}
	+ \frac{6}{7}\dW^{}_{a\langle i}\dI^{(4)}_{j\rangle a}
	- \frac{1}{7}\dY^{(3)}_{a\langle i}\dI^{}_{j\rangle a}
	- \dY^{}_{a\langle i}\dI^{(3)}_{j\rangle a}
	- 2\dX\,\dI^{(3)}_{ij}\nn\\
	& \qquad\qquad
	- \frac{5}{21}\dW^{(4)}_{a}\dI^{}_{ija}
	+ \frac{1}{63}\dW^{(3)}_{a}\dI^{(1)}_{ija}
	- \frac{25}{21}\dY^{(3)}_{a}\dI^{}_{ija}
	- \frac{22}{63}\dY^{(2)}_{a}\dI^{(1)}_{ija}
	+ \frac{5}{63}\dY^{(1)}_{a}\dI^{(2)}_{ija}\nn\\
	& \qquad\qquad
	+ 2\dW^{(3)}\dW^{}_{ij}
	+ 2\dW^{(2)}\dW^{(1)}_{ij}
	- \frac{4}{3}\dW_{\langle i}\dW^{(3)}_{j\rangle}
	+ 2\dW^{(2)}\dY^{}_{ij}
	- 4\dW_{\langle i}\dY^{(2)}_{j\rangle}\nn\\
	& \qquad\qquad
	+ \epsilon_{ab\langle i}\bigg[
	\frac{1}{3}\dI_{j\rangle a}\dZ^{(3)}_b
	- \dI_{j\rangle a}^{(3)}\dZ^{}_b
	+ \frac{4}{9}\dJ^{}_{j\rangle a}\dW^{(3)}_b
	- \frac{4}{9}\dJ^{}_{j\rangle a}\dY^{(2)}_b
	+ \frac{8}{9}\dJ^{(1)}_{j\rangle a}\dY^{(1)}_b\bigg]\Bigg\}+ \calO \left(\frac{1}{c^9}\right)\,.
\end{align}
For the other multipole moments (essentially the mass octupole and the current quadrupole), such relations are purely quadratic and consist of odd parity terms, and so are not affected by dimensional-regularization corrections. We do not report them here, as they can be found in Sec.~III.B of~\cite{FBI15}.

Since we have absorbed the correction $\delta_\text{indirect}\dM_{ij}$ into the redefinition of the canonical moment, the relation~\eqref{eq:resMij}, derived in the context of dimensional regularization, now holds with exactly the same definition for the source quadrupole moment as proposed in~\cite{MQ4PN_renorm}. Namely, $\dI_{ij}$ is the ``renormalized'' source quadrupole moment (in the sense of~\cite{MQ4PN_renorm}) which is given for quasi-circular orbits in Sec.~\ref{subsec:MQ4PN}, and which contains crucial contributions due to the (IR) dimensional regularization. Adding up the non-linear contributions described in Sec.~\ref{sec:nonlin}, we obtain the radiative moment measured at infinity. As we will see the above procedure yields the correct perturbative limit for compact binaries on circular orbits, in agreement with first-order black hole perturbation theory.

Interestingly, we found that the perturbative limit turns out also to be correct when using a simpler treatment of the IR divergences of the  source quadrupole moment, based on the Hadamard regularization in   ordinary three dimensions, rather than the dimensional regularization. In this case, the correct relation between canonical and source moments is given by Eq.~(1.1) of~\cite{MQ4PN_jauge}, instead of the $d$-dimensional result~\eqref{eq:resMij}. We find that the cubic interactions $\dM\times \dM\times \dW_{ij}$ and $\dM\times \dM\times \dY_{ij}$ present in three dimensions do contribute to the perturbative limit for circular orbits (note that $\dW$ is zero in this case), but in such a way as to cancel the contribution due to the difference between the Hadamard and dimensional regularization schemes for the mass quadrupole source moment. It turns out then that we also recover the correct perturbative limit. However, in contrast to the dimensional regularization, we do not expect the Hadamard regularization in three dimensions to lead to a well-defined and fully unambiguous result beyond this limit, so the above observation should be considered only as an interesting consistency check.

\section{Corrections to the source moments due to infrared commutators}
\label{sec:commutators}

In this section, we discuss the appearance of infrared (IR) ``d'Alembertian commutators'' in the expression of the PN-expanded near-zone metric, and their effects on the expression of the source moments. This feature starts at 4PN order and does not affect the 4PN equations of  motion, but should have been considered in previous works~\cite{MHLMFB20, MQ4PN_IR, MQ4PN_renorm}. We shall show that the IR commutators enter the flux and the $(2,2)$ mode for generic orbits, however they do not contribute in the quasi-circular limit. Since they are zero for quasi-circular orbits, all the results explicitly presented in the previous works~\cite{MHLMFB20, MQ4PN_IR, MQ4PN_renorm} are correct. Throughout this section, we will be based on the construction and computation of the source mass quadrupole, as described in~\cite{MHLMFB20}.

\subsection{Appearance of the commutators in the PN metric}\label{subsec:comm_appearance}

Crucial to the derivation of the source moments at a given PN order is the expression of the near-zone metric with the same precision. Hereafter, we will present the case of the source mass quadrupole at 4PN order, but the following discussion is easily generalized to any source moment. The gothic metric perturbation $h^{\mu\nu} \equiv \sqrt{-g} g^{\mu\nu} - \eta^{\mu\nu}$ obeys the gauge-fixed Einstein field equations
\begin{equation}\label{eq:comm_boxhmunu_gen}
	\Box h^{\mu\nu} = \frac{16\pi G}{c^4}\tau^{\mu\nu} \,,
\end{equation}
where the pseudo stress-energy tensor $\tau^{\mu\nu}$ is defined \emph{e.g.} in Eq.~(2.4) of~\cite{MHLMFB20}. As explained in~\cite{PB02}, the PN metric is obtained by solving iteratively the wave equation~\eqref{eq:comm_boxhmunu_gen} in the near zone, in such a way as to match the metric in the far zone. The constructed expansion of the gravitational field in the near zone, say $\bar{h}^{\mu\nu}$, is then given formally, to arbitrarily high orders, by~\cite{BBBFMc, MBBF17} 
\begin{align}\label{eq:hNZ}
	\bar{h}^{\mu\nu} &= \frac{16\pi G}{c^4} \biggl[ \text{FP} \BoxRm \left[ \bar{\tau}^{\mu\nu}\right] \biggr] + \sum_{\ell=0}^{+\infty}
	\sum_{j=0}^{+\infty} \Delta^{-j}\hat{x}_L \, \frac{1}{c^{2j}}\,f_L^{(2j)\mu\nu}(t) \, .
\end{align}
The second term on the right-hand side of~\eqref{eq:hNZ} is an homogeneous solution that is regular at the coordinate origin $r=0$. The explicit $d$-dimensional expressions of the time-dependent functions $f_L^{\mu\nu}(t)$, as functionals of the MPM expansion of $\tau^{\mu\nu}$ in the exterior zone, can be found in~\cite{MBBF17}.

Of interest to us here is the first term in~\eqref{eq:hNZ}. It is made by the PN-expanded retarded integral $\text{FP}\BoxRm$, \emph{i.e.}, the inverse d'Alembertian operator in which all retardations are formally expanded in a PN fashion, and regularized by means of the finite part procedure, $\text{FP}_{B=0}\int \dd^d \mathbf{x}'(|\mathbf{x}'|/r_0)^B[\cdots]$. Furthermore, the operator $\text{FP} \BoxRm$ acts on the near-zone expansion of the pseudo stress-energy tensor $\bar{\tau}^{\mu\nu}$, which can be derived at any PN order from the lower order expressions of $\bar{h}^{\mu\nu}$. The IR behavior of the integrals entering the propagator is naturally affected by the choice of regularization. In general, the PN expanded propagator in~\eqref{eq:hNZ} has no reason to commute with the d'Alembertian operator $\Box$. This is the root of the appearance of the IR commutators in the PN metric, that are defined as
\begin{equation}\label{comm_F_def}
	\comm \big\lbrace F\big\rbrace  \equiv \Bigl[\text{FP}\BoxRm,\Box\Bigr]F = \FP \overline{\Box}_\text{ret}^{-1}\biggl[\left(\frac{r}{r_0}\right)^B \Box F \biggr] - F\,,
\end{equation}
where $F$ is a regular (smooth) function in the near zone but typically diverges at spatial infinity, when $r\to+\infty$; see~\eqref{eq:comm_F_exp} below.

In order to simplify the resolution of the Einstein field equations~\eqref{eq:comm_boxhmunu_gen}, we parametrize the PN metric by means of some PN ``potentials'', obeying flat space-time wave equations in $d$ dimensions
\begin{equation}\label{BoxP}
	\Box P = S\,,
\end{equation}
where $S$ is composed of a compact-supported sector (proportional to Dirac distributions in the case of point particles), and a non-compact one (composed of products of derivatives of lower-order PN potentials). We refer to App.~A of~\cite{MHLMFB20} for complete definitions of the PN potentials. Now those potentials are constructed by means of the operator $\text{FP}\BoxRm$, \emph{i.e.}, they are given, by definition, as
\begin{equation}\label{P_def}
	P \equiv \text{FP} \BoxRm S\,,
\end{equation}
which accounts for the first term in the right side of~\eqref{eq:hNZ}. The second term in~\eqref{eq:hNZ} is non-local in time and corresponds to radiation modes that are being accounted for separately~\cite{BBBFMc}.  

When parametrizing the PN metric in this manner, it is very handy to recognize products of lower-order potentials. For instance, if some relevant combination of the field equations~\eqref{eq:comm_boxhmunu_gen} contains terms of the form $\partial_i P \partial_i P$ at some $n$PN order,\footnote{For convenience we pose $h^{00ii} \equiv \frac{2}{(d-1)} [(d-2)h^{00}+h^{ii}]$ in $d$ dimensions.}
\begin{equation}
	\Box \bar{h}^{00ii}_{(n)}  = \cdots + \partial_i P\partial_iP + \cdots\,,
\end{equation}
then it can be solved using the PN-expanded retarded propagator $\text{FP}\BoxRm$, as
\begin{align}\label{eq:comm_h_funcPS} 
	\bar{h}^{00ii}_{(n)}
	& \equiv \cdots + \text{FP}\BoxRm\big\lbrace\partial_i P\partial_i P\big\rbrace + \cdots
	= \cdots + \text{FP}\BoxRm\left\lbrace\frac{1}{2} \Box \left(P^2\right)- P \Box P\right\rbrace + \cdots \nonumber \\ &
	= \cdots + \frac{P^2}{2} - \text{FP}\BoxRm\big\lbrace P S\big\rbrace + \frac{1}{2} \comm\lbrace P^2\rbrace + \cdots \,,
\end{align}
where we have used Eq.~\eqref{BoxP} and neglected the higher PN order term $(\partial_t P)^2/c^2$ in the second curly brackets of the first line. Because solving $\text{FP}\BoxRm\lbrace PS\rbrace$ is much more tractable than the original $\text{FP}\BoxRm\lbrace \partial_iP\partial_iP\rbrace$, the operation described by Eq.~\eqref{eq:comm_h_funcPS} constitutes a real improvement. As shown later on, the commutator left in Eq.~\eqref{eq:comm_h_funcPS} is in general not vanishing. The expression of the 4PN metric in terms of those commutators is displayed in App.~\ref{app:comm_metric}.

While, formally, the commutators appear at a relative 1PN order in the metric [see Eqs.~\eqref{eq:app_comm_hmunu}], they effectively only contribute starting at 4PN order, as shown in Sec.~\ref{subsec:comm_MQ4PN}. Therefore, the only source multipole moment to be affected is the 4PN source mass quadrupole. Moreover, the 4PN contribution in the metric arises as a mere function of time in $\bar{g}_{00}$ only, through the combination $\bar{h}^{00ii}$. Since it is the spatial gradient $\partial_i \bar{g}_{00}$ that contributes to the equations of motion, the corresponding contribution vanishes. More generally, following the ``$n+2$ method''~\cite{BBBFMa} to determine the PN order of the metric needed for insertion in the Fokker Lagrangian, we see that the commutators play no role in the derivation of the conservative equations of motion at 4PN order.

\subsection{Expression of the commutators}\label{subsec:comm_expr_gen}

We immediately see from~\eqref{comm_F_def} that the commutator must be a homogeneous solution of the wave equation. Furthermore, for the IR commutator, which depends on the far zone behavior $r\to +\infty$, that homogeneous solution must be regular in the near zone, when $r\to 0$. These facts constrain its form: given the function $F$, there exist a set $\{\hat{g}_L\}$ of (yet unspecified) functions of time only, such that
\begin{equation}\label{eq:comm_F_gen}
	\comm \big\lbrace F\big\rbrace = \sum_{\ell= 0}^{+\infty}\frac{(-)^\ell}{\ell!}\sum_{j = 0}^{+\infty} \frac{1}{c^{2j}}\Delta^{-j}\hat{x}_L\, \hat{g}_L^{(2j)}(t)\,,
\end{equation}
where we recall that the inverse Laplacians are defined in Eq.~\eqref{notationLaplace}.
In order to explicitly determine the functions $\hat{g}_L$, we work out the propagator applied to a generic source $S(\mathbf{x},t)$:
\begin{align}\label{eq:comm_propF_gen}
	\text{FP} \BoxRm S =
	-\frac{\tilde{k}}{4\pi}\FP \sum_{k \geqslant 0}
	\int \!\!\dd^d \mathbf{x}' \left( \frac{r'}{r_0} \right)^B \frac{\vert \mathbf{x} - \mathbf{x}'\vert^{2k}}{2^{2k} k!\,c^{2k}}
	\biggl[
	&
	\frac{\Gamma\left(2-\frac{d}{2}\right)}{\Gamma\left(k+2-\frac{d}{2}\right)} \frac{S^{(2k)}(\mathbf{x}',t)}{\vert \mathbf{x} - \mathbf{x}'\vert^{d-2}}\nn\\
	&
	- \frac{\sqrt{\pi}}{2\Gamma\left(\frac{5-d}{2}\right) \Gamma\left(k+\frac{d}{2}\right)} \int_0^{+\infty}\!\! \dd \tau \,\frac{\tau^{3-d}}{c^{d-2}} S^{(2k+2)}(\mathbf{x}',t-\tau)
	\biggr]\,.
\end{align}
This expression comes from the near-zone expansion of the homogeneous solution of the wave equation in $d$ dimensions. The two terms in~\eqref{eq:comm_propF_gen} represent respectively the even and odd parity parts of this expansion, as given by Eqs.~(A7) and~(A8) in~\cite{BBBFMc}. Note that the odd parity part involves a non-local in time integral, shown in the second term of~\eqref{eq:comm_propF_gen}. The PN propagator~\eqref{eq:comm_propF_gen} corresponds exactly to the prescription which is required for the first term in Eq.~\eqref{eq:hNZ}, \emph{i.e.}, where all retardations of the inverse d'Alembertian operator are PN expanded. In three dimensions, this prescription recovers Eqs.~(3.4) and~(3.7) of~\cite{BFN05}. In turn, the prescription ensures the correct matching of the PN metric to the MPM metric in the exterior zone.

When computing the commutator $\mathcal{C}\lbrace F \rbrace$ with Eqs.~\eqref{comm_F_def} and~\eqref{eq:comm_propF_gen}, we may transform the d'Alembertian operator inside the source term by means of appropriate integrations by parts, which yields for the commutator a finite part integral with a global factor $B$ appearing in the source: 
\begin{equation}\label{comm_F_def_B}
	\comm \big\lbrace F\big\rbrace = \FP \overline{\Box}_\text{ret}^{-1}\biggl[ B \left(\frac{r}{r_0}\right)^B \left( - \frac{B-d}{r^2}F - \frac{2}{r}\partial_r F\right) \biggr] \,.
\end{equation}
Note the similarity with the expression of the quantities $X_n^{\mu\nu}$ and $Y_n^{\mu\nu}$ in~\eqref{eq:defXnYn}. Indeed, the latter have been formally defined as commutators (see Eq.~(3.20) of~\cite{MQ4PN_jauge}), but, in contrast to the ``IR'' commutators~\eqref{comm_F_def} or~\eqref{comm_F_def_B}, $X_n^{\mu\nu}$ and $Y_n^{\mu\nu}$ are ultraviolet (``UV'') commutators since they depend on the near zone behavior of the source term, when $r\to 0$.

We now use the expression~\eqref{comm_F_def_B} of the commutator. As we said, because of the explicit factor $B$ in the source, the commutator depends only on the IR behavior of the function $F$, \emph{i.e.}, when $r\to+\infty$ with $t=$ const (spatial infinity). We expand the function in $d=3+\varepsilon$ dimensions as
\begin{equation}\label{eq:comm_F_exp}
	F(\mathbf{x},t) = \sum_{p_0 \leqslant p\atop q_0 \leqslant q \leqslant q_1} \!\!\frac{f_{p,q}\left(\mathbf{n},t\right)}{r^{p+q \varepsilon}}\,,
\end{equation}
where the sums extend on all multipolarities $\ell$, and formally for all values of $p$ larger than some (generally negative) integer $p_0$, and for a finite set of values of $q$ for each given $p$. We inject this expansion into the commutator~\eqref{comm_F_def_B} and employ the explicit expression~\eqref{eq:comm_propF_gen} for the PN expanded propagator. Consistently with the IR limit we must expand the factors $|\mathbf{x} - \mathbf{x}'|^\alpha$ in~\eqref{eq:comm_propF_gen} (where $\alpha$ depends on the parity) when $|\mathbf{x}'|\to+\infty$. Actually this expansion is valid as soon as $r'=|\mathbf{x}'|>r=|\mathbf{x}|$ and reads, using the STF decomposition and the notation~\eqref{notationLaplace},\footnote{Eq.~\eqref{expinfty} is equivalent to the more familiar (but less convenient for our purpose) expansion in terms of Gegengauer polynomials $C_\ell^{\gamma}(x)$,
	$$\vert \mathbf{x}'-\mathbf{x}\vert^\alpha = {r'}^                   \alpha\sum_{\ell=0}^{+\infty} C_\ell^{-\frac{\alpha}{2}}\!(\mathbf{n}\cdot\mathbf{n}')\left(\frac{r}{{r}'}\right)^\ell \,.$$}
\begin{align}\label{expinfty}
	\vert \mathbf{x}'-\mathbf{x}\vert^\alpha = \sum_{\ell=0}^{+\infty} \frac{(-)^\ell}{\ell!} \sum_{j=0}^{+\infty} \Delta^{-j} \hat{x}_L \,\Delta'^j \hat{\partial}'_L r'^\alpha\,,
\end{align}
the object $\Delta'^j \hat{\partial}'_L r'^\alpha$ being explicitly known as
\begin{align}
	\Delta'^j \hat{\partial}'_L r'^\alpha = 2^{\ell+2j}\frac{\Gamma\bigl(\frac{\alpha}{2}+1\bigr)\Gamma\bigl(\frac{\alpha+d}{2}\bigr)}{\Gamma\bigl(\frac{\alpha}{2}+1-\ell-j\bigr)\Gamma\bigl(\frac{\alpha+d}{2}-j\bigr)}\hat{x}'_L r'^{\alpha-2\ell-2j}\,.
\end{align}
With the latter formulas at hand, it is straightforward to obtain the functions $\hat{g}_L(t)$ parametrizing the commutator in~\eqref{eq:comm_F_gen} in terms of the expansion coefficients in~\eqref{eq:comm_F_exp}. We find
\begin{align}\label{eq:comm_ghat_expr}
	\hat{g}_L(t)
	= \sum_{k= 0}^{+\infty} \frac{4k-2\ell-d+2}{2^{2k} k!\,c^{2k}(d-2)}\Bigg[
	&
	\frac{2^\ell\,\Gamma\left(2-\frac{d}{2}\right)}{\Gamma\left(k+2-\ell-\frac{d}{2}\right)} \,\hat{f}^{L\,(2k)}_{2k-\ell,0}(t) \\
	& 
	- \frac{\sqrt{\pi}}{2^{\ell+1}\Gamma\left(\frac{5-d}{2}\right) \Gamma\left(k+\ell+\frac{d}{2}\right)} \frac{4k+2\ell+d-2}{4k-2\ell-d+2}\int_0^{+\infty}\!\! \dd \tau \,\frac{(c \,\tau)^{3-d}}{c^{2\ell+1}} \,\hat{f}^{L\, (2k+2\ell+2)}_{2k+\ell+1,1}(t-\tau)
	\Bigg]\,,\nonumber
\end{align}
where we define the angular average over the $(d-1)$-dimensional sphere
\begin{equation}\label{sphereddim}
	\hat{f}^L_{p,q}(t) \equiv \int \frac{\dd\Omega_{d-1}}{\Omega_{d-1}}\, \hat{n}^L f_{p,q}(\mathbf{n},t)\quad~\text{with}~\quad\Omega_{d-1}=\frac{2\pi^{\frac{d}{2}}}{\Gamma(\frac{d}{2})}\,.	
\end{equation}
Note the non-local character of the relation~\eqref{eq:comm_ghat_expr} in $d$ dimensions, coming from the odd parity part of the PN retarded integral~\eqref{eq:comm_propF_gen}. Finally we see that only specific values of $q$ contribute: the $q=0$ terms induce an even sector, and the $q=1$ terms, an odd one.
Finally, it is important to remark that the formula~\eqref{eq:comm_ghat_expr} cannot induce poles $\propto (d-3)^{-1}$ by itself. Therefore, we can safely work in the $d=3$ limit, as long as the coefficients $\hat{f}^L_{p,q}$ do not have poles.

\subsection{Application to the source mass quadrupole}\label{subsec:comm_MQ4PN}

Let us apply the formulas~\eqref{eq:comm_F_gen} and~\eqref{eq:comm_ghat_expr} to the case of the source mass quadrupole at 4PN order (the IR commutators do not affect the computation of the other required moments). The fact that Eq.~\eqref{eq:comm_ghat_expr} selects only terms with $q=0$ or $q=1$ drastically reduces the number of interactions that effectively can play a role at 4PN. To lowest orders at least, one can show that the even PN sector of the potentials bears $q\geqslant 1$, and the odd one, $q=0$. Therefore, the only possible interactions at lowest orders are either even-odd or odd-odd couplings, as even-even ones will have $q\geqslant 2$. In addition, the odd part of both the lowest order potentials $V$ and $V_i$ (defined in the App.~A of~\cite{MHLMFB20}) starts at 1.5PN (more precisely, the 0.5PN term of those two potentials is vanishing in the $d \to 3$ limit). Investigating the metric~\eqref{eq:app_comm_hmunu}, one can see that only two interactions can lead to corrections with 4PN accuracy: $V^2$ and $V\hat{W}$, where we denote $\hat{W}\equiv\hat{W}_{kk}$.\footnote{The near-zone potential $\hat{W}_{ij}$ and its trace $\hat{W}$ (see App.~A of~\cite{MHLMFB20}) should not be confused with the $\ell$-th gauge moment $\dW_L$ and $\dW$ corresponding to $\ell=0$, as defined in~\eqref{eq:Jauge_vector}--\eqref{eq:Jauge_Ftilde_def}.} The commutators affect the quantities $\Sigma$ and $\tilde{\mu}_1$, defined in Eqs.~(2.10) and~(2.17) of~\cite{MHLMFB20}, as
\begin{subequations}
	\begin{align}
		&
		\Delta \Sigma =\frac{4(d-1)\sigma}{(d-2)^2c^4} \Bigg(
		\comm \big\lbrace V^2\big\rbrace 
		+   \frac{2(d-2)}{(d-1)c^2} \comm\big\lbrace V \hat{W}\big\rbrace \Bigg) + \calO\left(\frac{1}{c^{10}} \right)\,,\\
		&
		\Delta \tilde{\mu}_1 = \frac{2(d-4)}{(d-2)c^4} \Bigg(
		\comm \big\lbrace V^2\big\rbrace 
		+  \frac{2(d-2)}{(d-1)c^2} \comm\big\lbrace V \hat{W}\big\rbrace \Bigg)_{\!\! 1} m_1 + \calO\left(\frac{1}{c^{10}} \right)\,,
	\end{align}
\end{subequations}
where $\sigma$ is the compact-supported expression defined in Eq.~(2.15) of~\cite{MHLMFB20}, $m_1$ the mass of particle 1, and where the label 1 indicates that the quantities have to be evaluated and regularized at its location. This induces a correction to the source quadrupole
\begin{equation}\label{eq:comm_DeltaIij}
	\Delta \dI_{ij} = \frac{d(d-1)}{(d-2)^2c^4}\int\!\!\dd^d\mathbf{x} \Bigg(
	\comm \big\lbrace V^2\big\rbrace 
	+  \frac{2(d-2)}{(d-1)c^2} \comm\big\lbrace V \hat{W}\big\rbrace \Bigg)\,\delta_1 m_1 \hat{y}_1^{ij} + (1\leftrightarrow2) + \calO\left(\frac{1}{c^{10}} \right)\,,
\end{equation}
where $\delta_1 = \delta^{(d)}[\mathbf{x}-\bm{y}_1]$ is the $d$-dimensional Dirac distribution at the position $\bm{y}_1$ of the particle 1.

The commutator of the interaction $V^2$ is quite easy to compute. Indeed, the potential $V$ is known in the whole space to a high PN order, well beyond the required precision. One can straightforwardly expand it as in Eq.~\eqref{eq:comm_F_exp} and read the coefficients $\hat{f}_{p,q}^L$. As discussed previously, the 0.5PN term of $V$ vanishes in the $d=3$ limit. Thus, only the (even, 0PN)--(odd, 1.5PN) coupling, having $q=1$, can enter at 4PN. Working it explicitly, we find
\begin{equation}\label{eq:comm_commV2}
	\comm \big\lbrace V^2\big\rbrace = \frac{G^2 m}{c^4} \left[2 \tilde{\mu}^{(2)} c^2 + \frac{1}{3} I^{(4)} \right] + \mathcal{O}\left(\varepsilon\right) \,,
\end{equation}
where $m=m_1+m_2$ and $\tilde{\mu}(t) = \int \dd^3 \mathbf{x}'\, \sigma(\mathbf{x}',t)$ reduces to a constant at leading order for arbitrary matter systems, so that $\tilde{\mu}(t) = m + \mathcal{O}(c^{-2})$ with $m$ constant, while $I=\int \dd^3 \mathbf{x}'\, \sigma(\mathbf{x}',t) \,r'^2$ represents the effective moment of inertia. For compact binaries, those quantities read explicitly
\begin{align}
	\tilde{\mu}(t) &= \tilde{\mu}_1(t) + \tilde{\mu}_2(t)= m_1 \left[ 1 + \frac{1}{c^2} \left( \frac{3}{2} \bm{v}_1^2 - \frac{G m_2}{r_{12}} \right) \right] + (1 \leftrightarrow 2) + \mathcal{O}\left(\frac{1}{c^4}\right)\, ,\\
	I(t) &= m_1 \,\bm{y}_1^2 + m_2 \,\bm{y}_2^2 + \mathcal{O}\left(\frac{1}{c^2}\right) \, ,
\end{align}
with $\bm{v}_1=\dd \bm{y}_1/\dd t$ and $r_{12}=|\bm{y}_1 - \bm{y}_2|$. On the other hand, the $d$-dimensional expression of the potential $\hat{W}$ is not known in the whole space (see App.~C of~\cite{BDE04} for more details on the computation). Using an integration by part, one can rewrite the wave equation it obeys, Eq.~(A4d) of~\cite{MHLMFB20}, as
\begin{equation}
	\Box \hat{W} 
	= \frac{8\pi G}{d-2}\,\sigma_{kk} - \frac{d-1}{2(d-2)}\partial_iV\partial_i V
	= \frac{8\pi G}{d-2}\,\sigma_{kk} - \frac{d-1}{4(d-2)} \Big( \Box V^2 + \frac{2}{c^2} (\partial_t V)^2 + 8\pi G V \sigma \Big)\,,
\end{equation}
where $\sigma_{kk}$ and  $\sigma$ are the compact-supported expression defined in Eq.~(2.15) of~\cite{MHLMFB20}. The compact-supported part of the source is easily integrated over the whole space, using methods described \emph{e.g.} in~\cite{MHLMFB20}. It contains an even sector with $q=1$ that cannot contribute (we recall that, up to 1PN, $V$ bears $q=1$), and an odd, 0.5PN sector with $q=0$ that will contribute, when coupled to the expansion of $V$. As for the non-compact sector, it reads
\begin{equation}
	\hat{W}^\text{NC} 
	= - \frac{d-1}{4(d-2)} \text{FP}\BoxRm\bigg\lbrace \Box V^2 + \frac{2}{c^2} (\partial_t V)^2 \bigg\rbrace
	=- \frac{d-1}{4(d-2)} \bigg( V^2 + \frac{2}{c^2} (\partial_t V)^2 + \comm \big\lbrace V^2\big\rbrace\bigg) \,.
\end{equation}
The first two terms will have $q=2$ up to 1PN and 2PN orders, respectively, and we just proved that the third term is of order 2PN. Therefore, $\hat{W}^\text{NC}$ does not contribute to the IR commutator at the required order. Applying the formula~\eqref{eq:comm_ghat_expr}, one finds the nice relation
\begin{equation}\label{eq:comm_commVW}
	\comm \big\lbrace V\hat{W}\big\rbrace  = - \frac{3\,c^2}{4} \comm \big\lbrace V^2\big\rbrace + \calO\left(\varepsilon \right) + \calO\left(\frac{1}{c^4} \right)\,.
\end{equation}
With the two explicit expressions~\eqref{eq:comm_commV2} and~\eqref{eq:comm_commVW} at hand, one can compute the resulting correction to the source mass quadrupole $\Delta \dI_{ij}$ in~\eqref{eq:comm_DeltaIij}. In the CoM frame, it comes
\begin{equation}\label{eq:DeltaIij_comm}
	\Delta \dI_{ij} = -\frac{4G^3m^4\nu^2}{r^3\,c^8}\bigg[\bm{v}^2-3\dot{r}^2- \frac{Gm}{r}\bigg] x_{\langle i}x_{j \rangle} + \mathcal{O}\left(\frac{1}{c^{10}}\right)\,.
\end{equation}
In the case of quasi-circular orbits, $\bm{v}^2 = Gm/r + \calO(c^{-2})$ and $\dot{r} = \calO(c^{-5})$, so this correction plays no role. Note also that the two corrections to the metric~\eqref{eq:comm_commV2} and~\eqref{eq:comm_commVW} are functions of the time only, and so cannot contribute to the equations of motion.

\section{Source multipole moments for circular orbits}
\label{sec:source}

Having now expressed the radiative moments in terms of the canonical moments in Sec.~\ref{subsec:can2rad}, and the canonical moments in terms of the source and gauge moments in Sec.~\ref{subsec:source2can}, we now review the explicit expression of the source moments for compact binary systems. Note that the gauge moments (entering the relation~\eqref{eq:resMij} as well as in Sec.~III.B of~\cite{FBI15}) are required only at Newtonian order, excepted for $\dW$, needed at 1PN. They are easily computed using Eq.~(125) of~\cite{BlanchetLR}, and their expressions on quasi-circular orbits are displayed in Eq.~(5.7) of~\cite{FMBI12}. The most crucial moment is the source quadrupole moment $\dI_{ij}$, needed only at 4PN order. Indeed, its 4.5PN terms, due to 2PN relative radiation reaction corrections, cannot involve non-local integrals, and will disappear from the flux.

\subsection{Source mass quadrupole at the 4PN order}\label{subsec:MQ4PN}

A first computation, using the Hadamard regularization for the IR divergences and dimensional regularization for the UV ones, was performed in~\cite{MHLMFB20}. However, in order to be consistent with the equations of motion~\cite{BBBFMa,BBBFMb,BBBFMc,BBFM17} which used dimensional regularization both in the UV and IR sectors, it should be computed with dimensional regularization also for the IR divergences. It was thus completed in~\cite{MQ4PN_IR}, and the expression of the source quadrupole moment, obtained with full dimensional regularization, has the expected feature of exhibiting poles in $\varepsilon = d-3$. As reviewed in Sec.~\ref{subsec:renorm}, those poles are crucial to cancel the divergences linked with the $d$ dimensional computation of the radiative moments, obtained in~\cite{MQ4PN_renorm}. Indeed, the source moments are not observables \emph{per se}, but the radiative moments are. Therefore, only the latter have to be finite in the $\varepsilon \rightarrow 0$ limit. However, as already mentioned, for the sake of computational simplicity, it was deemed possible to introduce the notion of a ``renormalized'' source quadrupole, defined by Eq.~(6.2) of~\cite{MQ4PN_renorm}. This renormalized quantity is nothing but the sum of the $d$-dimensional source quadrupole and the corrections~\eqref{eq:dUij_renorm}, arising from the radiative/canonical relation (we recall that we consider the corrections~\eqref{eq:gauge_dMij_ddim2} as being part of the canonical/source relation).
The renormalized quadrupole, by construction, has no poles in $\varepsilon$, and when injected into the three dimensional MPM algorithm, yields the correct radiative quadrupole by definition. 
Last, but not least, up to 3.5PN order, the corrections due to the IR dimensional regularization exactly cancel the ones due to the renormalization, even out of the CoM frame, as proven in~\cite{MQ4PN_renorm}. Thus, up to 3.5PN order, the renormalized source quadrupole coincides with the one computed with Hadamard regularization in the IR, which is why those subtleties did not hit previous computations of the flux. However, this equivalence is no more valid at 4PN order, thus the crucial need for the implementation of dimensional regularization and the ``renormalization'' described above.

As the correction due to the IR commutators discussed in Sec.~\ref{sec:commutators} vanishes on quasi-circular orbits, the expression of the renormalized source quadrupole on quasi-circular orbits remains identical to the one displayed in Eq.~(6.11) of~\cite{MQ4PN_renorm}, namely
\begin{equation}\label{eq:Iijcirc}
\dI_{ij} = m\nu\, \left(
	A \, x_{\langle i}x_{j \rangle}
	+B \, \frac{r^2}{c^2}v_{\langle i}v_{j \rangle}
	+ \frac{G^2 m^2\nu}{c^5r}\,C\,x_{\langle i}v_{j \rangle}\right)
	+ \mathcal{O}\left(\frac{1}{c^{9}}\right)\,,
\end{equation}
where, introducing the PN parameter $\gamma \equiv \frac{Gm}{rc^2}$, the coefficients are given by
\begin{subequations}\label{eq:Iijrenorm_AB}
	\begin{align}
		A =\ & 1
		+ \gamma \biggl(- \frac{1}{42}
		-  \frac{13}{14} \nu \biggr)
		+ \gamma^2 \biggl(- \frac{461}{1512}
		-  \frac{18395}{1512} \nu
		-  \frac{241}{1512} \nu^2\biggr)
		\nn\\
		& + \gamma^3 \biggl(\frac{395899}{13200}
		-  \frac{428}{105} \ln\biggl(\frac{r}{r_{0}{}} \biggr)
		+ \biggl[\frac{3304319}{166320}
		-  \frac{44}{3} \ln\biggl(\frac{r}{r'_{0}}\biggr) \biggr]\nu
		+ \frac{162539}{16632} \nu^2 + \frac{2351}{33264} \nu^3
		\biggr)
		\nn\\
		& + \gamma^4 \biggl (- \frac{1067041075909}{12713500800}
		+ \frac{31886}{2205} \ln\biggl(\frac{r}{r_{0}{}} \biggr)\nn\\
		& \qquad\quad
		+ \biggl[-\frac{85244498897}{470870400}
		-  \frac{2783}{1792} \pi^2-\frac{64}{7}\ln\left(16\gamma \de^{2\gamma_\text{E}}\right)  -  \frac{10886}{735} \ln\biggl(\frac{r}{r_{0}{}} \biggr)
		+ \frac{8495}{63} \ln\biggl(\frac{r}{r'_{0}} \biggr)\biggr] \nu\nn\\
		& \qquad\quad
		+ \biggl[\frac{171906563}{4484480} + \frac{44909}{2688} \pi^2-  \frac{4897}{21}
		\ln\biggl(\frac{r}{r'_{0}} \biggr)\biggr]\nu^2 - \frac{22063949}{5189184} \nu^3 + \frac{71131}{314496} \nu^4
		\biggl)\,, \\
		%%%%%%%%%%%%%%%%%%%%%%%%%%%%%%%%%%%%%%%%%%%%%%%%%%%%%%%%%%%%%%%%%%%%%
		B =\ & \frac{11}{21}
		-  \frac{11}{7} \nu
		+ \gamma \biggl(\frac{1607}{378}
		-  \frac{1681}{378} \nu
		+ \frac{229}{378} \nu^2\biggr) \nn\\
		& + \gamma^2 \biggl(- \frac{357761}{19800}
		+ \frac{428}{105} \ln\biggl(\frac{r}{r_{0}{}} \biggr)
		-  \frac{92339}{5544} \nu
		+ \frac{35759}{924} \nu^2
		+ \frac{457}{5544} \nu^3 \biggr)  \nn\\
		& + \gamma^3 \biggl(\frac{23006898527}{1589187600} -  \frac{4922}{2205}
		\ln\biggl(\frac{r}{r_{0}{}} \biggr)\nn\\
		& \qquad\quad + \biggl[\frac{8431514969}{529729200}
		+ \frac{143}{192} \pi^2-\frac{32}{7}\ln\left(16\gamma \de^{2\gamma_\text{E}}\right) 
		-  \frac{1266}{49} \ln\biggl(\frac{r}{r_{0}{}} \biggr)
		- \frac{968}{63} \ln\biggl(\frac{r}{r'_{0}} \biggr)\biggr] \nu  \nn\\
		& \qquad\quad + 
		\biggl[\frac{351838141}{5045040} 
		-  \frac{41}{24} \pi^2
		+ \frac{968}{21} \ln\biggl(\frac{r}{r'_{0}} \biggr)\biggr] \nu^2
		-  \frac{1774615}{81081} \nu^3
		-  \frac{3053}{432432} \nu^4 \biggl)\,, \\
		%%%%%%%%%%%%%%%%%%%%%%%%%%%%%%%%%%%%%%%%%%%%%%%%%%%%%%%%%%%%%%%%%%%%%%
		C =\ & \frac{48}{7} + \gamma \left(-\frac{4096}{315} - \frac{24512}{945}\nu \right)-\frac{32}{7}\pi\,\gamma^{3/2}\,.
	\end{align}
\end{subequations}
The coefficients $A$ and $B$ represent the conservative part of the quadrupole, while $C$ is due to the radiation reaction dissipative effects. Note that, in addition to the already-encountered scale $r_0$, the expression of the quadrupole involves a scale $r_0'$, associated with the UV regularization (see the footnote 10 of~\cite{MHLMFB20} for more details). 

As discussed in~\cite{MQ4PN_IR}, we found that the source quadrupole moment is not a local quantity at the 4PN order anymore, as it contains a non-local tail integral, given by Eq.~(6.5) in~\cite{MQ4PN_renorm}. The 4PN logarithms $\ln(16\gamma \,\de^{2\gamma_\text{E}})$ in $A$ and $B$ are due to the conservative part of this non-local tail term in the mass quadrupole, and the coefficient $-\frac{32}{7}\pi \gamma^{3/2}$ in $C$, to the corresponding dissipative part of the tail term.

Finally, as reported in App.~\ref{app:BSS}, the general expression of the source mass quadrupole moment for generic orbits has been conclusively tested in the so-called boosted Schwarzschild black hole limit.

\subsection{Higher-order source moments}

As expected from the behavior of the associated radiative moment~\eqref{eq:dUijk_renorm}, poles arise when performing the IR dimensional regularization of the source mass octupole $\dI_{ijk}$. Nevertheless, the renormalized source mass octupole moment exactly coincides at 3PN order with the one computed with Hadamard regularization in the IR, see~\cite{MQ4PN_renorm} for discussion. It reads~\cite{FBI15}
\begin{equation}\label{Iijk}
	\dI_{ijk} = - \nu\,m\,\Delta\biggl\{D\,x_{\langle i}x_{j}x_{k\rangle} +
	E\,\frac{r}{c}\,v_{\langle i}x_{j}x_{k\rangle} +
	F\,\frac{r^2}{c^2}\,v_{\langle i}v_{j}x_{k\rangle} +
	G\,\frac{r^3}{c^3}\,v_{\langle i}v_{j}v_{k\rangle} \biggr\} + \mathcal{O}\left(\frac{1}{c^{7}}\right)\,,
\end{equation}
where $\Delta \equiv (m_1-m_2)/m$, and the coefficients are
\begin{subequations}\label{Iijk_DEFG}
	\begin{align}
		D =\ & 1 -\gamma \nu + \gamma^2 \left( - \frac{139}{330} -
		\frac{11923}{660}\nu - \frac{29}{110}\nu^2\right) \nn \\
		 & +
		\gamma^3 \left( \frac{1229440}{63063} + \frac{610499}{20020}\nu +
		\frac{319823}{17160} \nu^2 - \frac{101}{2340} \nu^3 - \frac{26}{7} \ln
		\Bigl(\frac{r}{r_0}\Bigr) - 22 \nu \ln \Bigl(\frac{r}{r'_{0}}\Bigr)\right)\,,\\
		%%%%%%%%%%%%%%%%%%%%%%%%%%%%%%%%%%%%%%%%%%%%%%%%%%%%%%%%%%%%%%%%%
		E =\ & \frac{196}{15}\gamma^2 \nu \,,\\
		%%%%%%%%%%%%%%%%%%%%%%%%%%%%%%%%%%%%%%%%%%%%%%%%%%%%%%%%%%%%%%%%%%
		F =\ & 1 - 2\nu +
		\gamma \left(\frac{1066}{165} - \frac{1433}{330}\nu + \frac{21}{55}\nu^2\right) + \gamma^2 \left( - \frac{1130201}{48510} - \frac{989}{33} \nu + \frac{20359}{330} \nu^2 - \frac{37}{198} \nu^3 + \frac{52}{7} \ln \Bigl(\frac{r}{r_0}\Bigr)\right)\,,\\
		%%%%%%%%%%%%%%%%%%%%%%%%%%%%%%%%%%%%%%%%%%%%%%%%%%%%%%%%%%%%%%%%%%
		G =\ & 0\,.
	\end{align}
\end{subequations}
As for the renormalized current quadrupole $\dJ_{ij}$, it also coincides at 3PN order with the one computed with Hadamard regularization in the IR~\cite{MQ4PN_renorm}. It comes~\cite{HFB_courant}
\begin{equation}\label{J2circ}
\dJ_{ij} = - \nu m \Delta\left[ H \,L^{\langle i} x^{j\rangle} + K \,\frac{G m}{c^3}\,L^{\langle i} v^{j\rangle} \right] + \mathcal{O}\left(\frac{1}{c^7}\right)\,,
\end{equation}
where we denote $L^i \equiv \epsilon_{ijk}x^jv^k$, and where
\begin{subequations}\label{J2circ_HK}
	\begin{align}
		H =\ & 1 +\gamma \left(\frac{67}{28}-\frac{2}{7}\nu \right)+\gamma^2\left(\frac{13}{9} -\frac{4651}{252}\nu -\frac{\nu^2}{168} \right) \nn\\
		& + \gamma^3\left(\frac{2301023}{415800} - \frac{214}{105}\ln\left(\frac{r}{r_0}\right) + \left[ - \frac{243853}{9240} + \frac{123}{128}\pi^2 - 22 \ln\left(\frac{r}{r'_0}\right)\right]\nu + \frac{44995}{5544}\nu^2 + \frac{599}{16632}\nu^3\right)\,,\\
		K =\ & \frac{188}{35} \nu \,\gamma\,.
\end{align}\end{subequations}
The remaining required source moments (mass hexadecapole at 2PN, current octupole at 2PN, \emph{etc.}) are easy to compute as no regularization subtleties arise. Their expressions on quasi-circular orbits are displayed \emph{e.g.} in Sec.~9.1 of~\cite{BlanchetLR}. Note that, as the first non-local feature cannot appear at a lower order than 4PN, all those higher-order source moments are instantaneous up to 3.5PN order. Therefore, they cannot contribute to the 4.5PN term of the quasi-circular flux.

\section{Toolbox of integration formulas}\label{sec:toolbox}

This technical section presents some miscellaneous material that was used to perform the explicit computations of the gravitational wave flux to 4.5PN and the $(2,2)$ mode to 4PN, which will be reported in Sec.~\ref{sec:res}.

\subsection{4PN equations of motion for circular orbits}

First, the relation between the radiative and canonical moments, as well as the expression of the flux, involve temporal derivatives. We thus need the expression of the equations of motion for quasi-circular orbits at the 4PN order, including both conservative and dissipative terms. Note that the 4.5PN piece of the equations of motion~\cite{GII97,Leibovich:2023xpg} is local and so, as already discussed, will not contribute to the 4.5PN flux. Recalling that $\gamma=\frac{G m}{r c^2}$, the acceleration for quasi-circular orbit reads~\cite{BBFM17}
\begin{equation}\label{eomcirc}
\frac{\dd v^i}{\dd t} = -\omega^2 x^i - \frac{32}{5}\frac{G^3m^3\nu}{r^4\,c^5}\left[1 +\left(-\frac{743}{336} - \frac{11}{4}\nu\right)\gamma + 4\pi\gamma^{3/2} + \mathcal{O}\bigl(\gamma^2\bigr)\right] v^i \,.
\end{equation}
The radiation-reaction part (proportional to $v^i$) includes 1PN and 1.5PN corrections beyond the leading 2.5PN effect. In particular the 1.5PN correction is due to tails. The conservative part of the equations of motion is specified for circular orbits by the orbital frequency, which is a generalization of Kepler's law valid through 4PN:
\begin{align}\label{keplerlaw}
	\omega^2 &= \frac{G m}{r^3} \Bigg\{ 1+(-3+\nu) \gamma + \left( 6 +
	\frac{41}{4}\nu + \nu^2 \right) \gamma^2 \nn\\ & \quad\quad +
	\Biggl( -10 + \left[- \frac{75707}{840} + \frac{41}{64} \pi^2 + 22
	\ln \left( \frac{r}{r'_0}\right) \right]\nu + \frac{19}{2}\nu^2 +
	\nu^3 \Biggr) \gamma^3 \nn\\ & \quad\quad +
	\Biggl(15 + \left[\frac{19644217}{33600} + \frac{163}{1024} \pi^2
	+ \frac{256}{5} \gamma_\text{E} + \frac{128}{5} \ln (16 \gamma ) 
	- 290 \ln\Big(\frac{r}{r'_{0}}\Big) \right] \nu \nn\\
	&\qquad\qquad\quad  + \left[\frac{44329}{336} 
	-  \frac{1907}{64} \pi^2 + \frac{1168}{3} \ln\Big(\frac{r}{r'_{0}}\Big) \right] \nu^2 + \frac{51}{4} \nu^3 
	+ \nu^4 \Biggr)\gamma^4 + \mathcal{O}\bigl(\gamma^5\bigr)\Biggr\}\,.
\end{align}
We recall that the scales $r_0$ and $r_0'$ are respectively associated with IR and UV regularizations. Using the flux-balance equation as well as the generalized Kepler's law, we find the secular decay of orbital quantities at 1.5PN order
\begin{subequations}\label{eq:rdot_omegadot}
\begin{align}
\dot{r} \equiv \frac{\dd r}{\dd t}  
& = - \frac{64}{5} \frac{G^3m^3\nu}{r^3\,c^5} \left[1 +\left(-\frac{1751}{336} - \frac{7}{4}\nu\right)\gamma + 4\pi	\gamma^{3/2} + \calO\bigl(\gamma^2\bigr)\right]\,,\\
\dot{\omega} \equiv \frac{\dd \omega}{\dd t} 
& = \frac{96}{5}\frac{Gm\nu}{r^3} \gamma^{5/2}\left[1 +\left(-\frac{2591}{336} - \frac{11}{12}\nu\right)\gamma + 4\pi \gamma^{3/2} + \calO\bigl(\gamma^2\bigr)\right]\,.\label{eq:omegadot}
\end{align}
\end{subequations}
The relative 2PN order is also known~\cite{GII97,Leibovich:2023xpg}, but is not required for the present work. In the following, it will be important to distinguish between the orbital frequency $\omega$ defined from the equations of motion by~\eqref{keplerlaw}, and the GW half-frequency $\Omega$ which is measured in the waves propagating in the far zone. We pose
\begin{equation}\label{eq:y_def}
	y \equiv \left(\frac{Gm\omega}{c^3}\right)^{2/3}\,,
\end{equation}
and distinguish it later from the PN parameter $x$ defined from the GW half-frequency $\Omega$, see Eq.~\eqref{eq:x_def}.

\subsection{Post-adiabatic integration of the tail effect}
\label{sec:PA}

In order to compute the tail integrals, for instance~\eqref{Uij15PN}, one needs to specify the orbit's behavior of the compact binary system in the remote past, as the effect is not localized in time, but integrates over the whole past history of the source. In the case of quasi-circular orbits, and up to 3.5PN precision in the multipoles, an adiabatic approximation (considering the orbital elements $r$ and $\omega$ to be constant in time) is sufficient, and one can follow the lines of~\cite{ABIQ04}, together with the integrals presented in the App.~B of~\cite{BS93}. However, Eqs.~\eqref{eq:rdot_omegadot} show that this adiabatic approximation is no longer valid at a relative 2.5PN precision. As the tail enters at 1.5PN order in the moments, the first ``post-adiabatic'' (PA) correction will affect the moments at the 4PN order, thus we need to properly evaluate it in order to consistently derive the 4PN flux and $(2,2)$ mode.

As shown in~\cite{ABIQ04}, the tail integrals on quasi-circular orbits reduce to elementary integrals of the type
\begin{align}\label{Ian}
\mathcal{I}_{a,n}(u) = \int_0^{+\infty} \dd\tau \, 
\bigl[y(u-\tau)\bigr]^a \,\de^{-\di n \,\phi(u-\tau)} \ln \left(\frac{\tau}{\tau_0}\right) \,,
\end{align}
where $n \in \mathbb{N}^\star$ (the case $n=0$ does not appear at 4PN), $a$ is usually a rational fraction, and the constant $\tau_0$ denotes either $2r_0/c$ or $2b_0/c$. The orbital phase $\phi$, the orbital frequency $\omega=\dot{\phi}$ and the parameter $y=(\frac{G m\omega}{c^3})^{2/3}$ are integrated over any instant $u-\tau$ in the past. Our strategy will be to notice that the integral $\mathcal{I}_{a,n}$ involves a fast oscillating exponential, that is derivable under properly defined PA approximations. This section provides a general method valid an arbitrary PA order, which we apply to first order 1PA. 

We first remark that the difference between the current phase $\phi(u)$ and the phase $\phi(u-\tau)$ in the past is of the order of the inverse of the radiation reaction scale, \emph{i.e.}, $\phi(u) - \phi(u-\tau)$ scales as $\mathcal{O}(c^5)$, which is also obvious from Eq.~(317) of~\cite{BlanchetLR}. To describe the radiation-reaction scale, we introduce a dimensionless adiabatic parameter at the current time $u$, denoted by\footnote{This definition is intended to be valid at any PN order, and is different from the one adopted in previous work; see (4.8) in~\cite{ABIQ04}.}
\begin{align}
	\xi(u) \equiv \frac{\dot{\omega}(u)}{\omega^2(u)} = \mathcal{O}\left(\frac{1}{c^5}\right)\,.
\end{align}	
Using the energy balance equation we have $\xi=-\mathcal{F}(\omega)/[\omega^2 \dd E/\dd \omega]$. We want to compute the integral~\eqref{Ian} in the PA limit where $\xi(u)\to 0$. To this end, let us pose
\begin{align}\label{vdef}
	\phi(u) - \phi(u-\tau) = \frac{v}{\xi(u)}\,,
\end{align}
and change the integration variable from $\tau$ to $v$, so that the integral~\eqref{Ian} becomes
\begin{align}\label{Ian2}
	\mathcal{I}_{a,n} =  \frac{Gm}{c^3}\,\frac{\de^{-\di n \phi(u)}}{\xi(u)}\,\int_0^{+\infty} \dd v\, 
	\bigl[y(u-\tau(v))\bigr]^{a-3/2}  \,\ln \Bigl(\frac{\tau(v)}{\tau_0}\Bigr) \,\de^\frac{\di n \,v}{\xi(u)}\,.
\end{align}
Here $\tau(v)$ is obtained by inverting Eq.~\eqref{vdef} for $\tau$ as a function of $v$. The main point is that, in the PA limit, $\xi(u)\to 0$, and so, as the imaginary exponential in the integrand of~\eqref{Ian2} oscillates very rapidly, the integral is a sum of alternatively positive and negative contributions which essentially sum up to zero. However there are no oscillations when $v=0$, therefore the integral is essentially given by the contribution in the neighborhood of the bound at $v=0$. In other words we are entitled to compute the integral by replacing the integrand by its formal expansion series when $v\to 0$. This will give a formal (asymptotic) series in powers of $\xi(u)$ which is the ``PA'' expansion of the integral.

In the PA limit we thus have $\tau\to 0$, and we can expand Eq.~\eqref{vdef} to any order as
\begin{align}\label{vdefsol}
	v = \xi \sum_{n=0}^{+\infty} \frac{(-)^n}{(n+1)!} \,\omega^{(n)} \tau^{n+1}\,.
\end{align}
From now on, we pose $\xi=\xi(u)$, $\phi=\phi(u)$, $\omega=\omega(u)=\dot{\phi}(u)$, $\omega^{(n)}=(\dd^n\omega/\dd u^n) (u)$ for the quantities at the \emph{current} time $u$. Since $\omega^{(1)}\equiv\dot{\omega}=\xi\,\omega^2$ the expansion~\eqref{vdefsol} clearly represents the PA approximation in powers of $\xi$ and arbitrary high time derivatives of $\xi$ denoted $\xi^{(n)}$. Note that each time derivative of $\xi$ adds a radiation reaction scale, hence we have $\xi^{(n)} = \mathcal{O}(\xi^{n+1}) = \mathcal{O}(c^{-5n-5})$. 

To obtain the PA expansion of~\eqref{Ian2} we need to invert the series~\eqref{vdefsol} and obtain $\tau$ as a power series in $v$. This is given by the Lagrange inversion theorem as
\begin{align}
	\tau = \sum_{p=0}^{+\infty} \frac{1}{(p+1)!}\,f_p\left(\frac{v}{\xi}\right)^{p+1}\,,
\end{align}
where the general coefficients $f_p$ read
\begin{align}\label{taudefsol}
	f_p = \left(\frac{\dd}{\dd\tau}\right)^p\Biggl[\left(\frac{\tau}{\phi(u)-\phi(u-\tau)}\right)^{p+1}\Biggr]_{\tau=0} = \left(\frac{\dd}{\dd\tau}\right)^p\Biggl[\left(
	\sum_{n=0}^{+\infty} \frac{(-)^n}{(n+1)!} \,\omega^{(n)} \tau^{n}
	\right)^{\!-p-1\,}\Biggr]_{\tau=0}\,,
\end{align}
where $\tau=0$ is applied after the $p$ differentiations with respect to $\tau$. For instance, up to 2PA order, we obtain
\begin{align}\label{tau2PA}
	\tau = \frac{v}{\xi\,\omega}\Biggl[1 + \frac{v}{2} + \biggl(1 - \frac{\dot{\xi}}{\xi^2 \omega}\biggr)\frac{v^2}{6} + \mathcal{O}\left(v^3\right)\Biggr]\,.
\end{align}

The previous formulas show that the method can be straightforwardly extended to any order. But as we said, to compute the tail term at 4PN order, we need only the correction of order 2.5PN, which corresponds to the 1PA approximation, hence just $\tau = \frac{v}{\xi\,\omega}[1 + \frac{v}{2} + \mathcal{O}(v^2)]$. To 1PA order, the integral~\eqref{Ian2} reads
\begin{align}\label{Ian3}
	\mathcal{I}_{a,n} =  \frac{Gm}{c^3}\,\frac{y^{a-3/2}\,\de^{-\di n \phi}}{\xi}\,\int_0^{+\infty} \dd v\, 
	\Biggl[ \biggl( 1+ \Bigl(1-\frac{2a}{3}\Bigr)v \biggr)\ln\left(\frac{v}{\xi\,\omega\,\tau_0}\right) + \frac{v}{2} + \mathcal{O}\left(v^2\right)\Biggr] \,\de^\frac{\di n \,v}{\xi}  \,.
\end{align}
The remaining integral is computed as follows. We transform the complex exponential into a real one by performing the change of variable $v=\di\, w$ for $n> 0$, and $v=-\di\, w$ for $n< 0$, respectively. The integration now takes place along the imaginary axis, which can be remedied by resorting to the Cauchy theorem on the closed contour made of the three following pieces, to be considered in the limit $R\to +\infty$: (i) the path from $w=0$ to $w=R$ on the real axis, (ii)~the oriented quarter of circle of radius $|w|=R$ from $\arg w = 0$ to $\arg w=-\pi/2$ (or $\arg w=\pi/2$ if $n < 0$) and (iii)~the segment of the imaginary axis going from $w=-\di R$ ($w=\di R$ if $n<0$) to $w=0$. This leads to
\begin{align}\label{Ian4}
	\mathcal{I}_{a,n} =  \frac{Gm}{c^3}\,\frac{y^{a-3/2}\,\de^{-\di n \phi}}{\di\,\xi}\,\int_0^{+\infty} \dd w\, 
	\Biggl[ \biggl( - \text{s}(n)+ \frac{2a-3}{3}\,\di\,w \biggr)\left(\ln\left(\frac{w}{\xi\,\omega\,\tau_0}\right) +\di \frac{\pi}{2} \,\text{s}(n)\right) + \frac{\di \,w}{2} + \mathcal{O}\left(w^2\right)\Biggr] \,\de^{-\frac{\vert n\vert \,w}{\xi}}  \,,
\end{align}
where $\text{s}(n)$ is the sign function. In this form, $\mathcal{I}_{a,n}$ can be integrated, as it boils down to elementary integrals. For completeness we give the formulas needed to handle any PA approximation:
\begin{subequations}
	\begin{align}
		\int_0^{+\infty} \dd w\, w^j \,\de^{-\frac{\vert n\vert \,w}{\xi}}
		&
		= j! \left(\frac{\xi}{\vert n \vert}\right)^{j+1}\,, \\ 
		\int_0^{+\infty} \dd w\, w^j \, \ln\left(\frac{w}{\xi\,\omega\,\tau_0}\right) \,\de^{-\frac{\vert n\vert \,w}{\xi}}	&
		= j! \left(\frac{\xi}{\vert n \vert}\right)^{j+1} \Big[ H_j - \gamma_\text{E} - \ln \bigl(\vert n \vert \,\omega \,\tau_0\bigr) \Big]\,,
	\end{align}
\end{subequations}
where $H_j=\sum_{k=1}^j \frac{1}{k}$ denotes the harmonic number and $\gamma_\text{E}$ is the Euler constant. Finally we obtain, at desired 1PA order, the result
\begin{equation}
	\mathcal{I}_{a,n} = \frac{Gm}{c^3}\,\frac{y^{a-3/2}\,\de^{-\di n \phi}}{\di \,n}\Bigg\lbrace \bigg( 1 + \frac{2a-3}{3}\,\frac{\xi}{\di \,n}\bigg)\bigg[ \ln\bigl(\vert n \vert \,\omega \,\tau_0\bigr) + \gamma_\text{E} - \di \,\frac{\pi}{2} \,\text{s}(n) \bigg] - \frac{4a-9}{6}\,\frac{\xi}{\di\,n} + \calO\left(\xi^2\right)\Bigg\rbrace\,.
\end{equation}
Recall that all quantities are evaluated at current time $u$ and that the adiabatic parameter $\xi = \dot{\omega}/\omega^2$ is easily computed with Eqs.~\eqref{eq:rdot_omegadot}. With this result at hand, we are able to derive the tail integral~\eqref{Uij15PN} with 2.5PN relative precision, which impacts the computation and final results at the 4PN order.

\subsection{Memory effects for circular orbits up to 1.5PN relative order}\label{subsec:memoryInt}

Memory terms, such as the first term of Eq.~\eqref{Uij25PN}, are hereditary integrals of the form $\int_0^{+\infty}\dd \tau F(u-\tau) G(u-\tau)$, where $F$ and $G$ represent dynamical multipole moments. Note that they enter only in mass-type moments, as clear from Sec.~\ref{subsec:can2rad}. In the case of quasi-circular orbits, they reduce to a sum of terms of the form
\begin{align}\label{eq:mem_Jan}
	\mathcal{J}_{a,n}(u) = \int_0^{+\infty} \dd\tau \, 
	\bigl[y(u-\tau)\bigr]^a \,\de^{-\di n \,\phi(u-\tau)} \,.
\end{align}
Besides the absence of logarithm, the main difference with the tail integral is the possibility of having $n = 0$, \emph{i.e.} persistent or ``DC'' terms.

Let us first focus on oscillatory (``AC'') memory terms, having $n \neq 0$. As they only involve a simple, logarithmic-free, integration, memory terms will enter in the flux as instantaneous contributions, and can be computed as such. In particular, since they arise at the odd PN approximations 2.5PN and 3.5PN, they do not contribute to the flux for quasi-circular orbits. The evaluation of the integrals of type~\eqref{eq:mem_Jan} is thus only required for the derivation of the modes. Concerning the quadrupole moment, since the memory effect enters at 2.5PN order, as clear from~\eqref{Uij25PN}, it is thus required at a relative 1.5PN precision for the mode $(2,2)$ and thus, one can safely compute the integral in the adiabatic approximation. As discussed in~\cite{ABIQ04}, for circular orbits this is equivalent to taking $y$ to be constant together with a linear phase (appropriate for the exact circular orbit), and keeping only the contribution of the integral due to the bound $\tau=0$. One finds
\begin{equation}
	\mathcal{J}_{a,n}  =  y^a \,\de^{-\di n \,\phi} \int_0 \dd\tau \,\de^{\di n \,\omega\,\tau} + \calO\left(\xi\right) = - \frac{y^a \,\de^{-\di n \,\phi}}{\di n \,\omega} + \calO\left(\xi\right) \,.
\end{equation}

The persistent DC terms obviously do not contribute to the flux, and they only contribute to the modes which have $\text{m}=0$, for example, the $(\ell,\text{m}) = (2,0)$  mode for the mass quadrupole. As is clear in the following, the absence of the fast oscillating exponential in $\mathcal{J}_{a,0}$ generates an inverse power of the adiabatic parameter $\xi$, thus degrading the precision by the radiation-reaction scale 2.5PN. This is the well-known memory effect: as it starts at 2.5PN order in the waveform, it finally enters the $(2,0)$ mode at Newtonian order. Several methods are possible to evaluate $\mathcal{J}_{a,0}$; see \emph{e.g.}~\cite{ABIQ04}. In the following, we rely on a change of integration variables from $\tau$ to $y'=y(u-\tau)$: 
\begin{align}\label{eq:Ja0}
	\mathcal{J}_{a,0} 
	= \int_0^{+\infty} \dd\tau \, 
	\bigl[y(u-\tau)\bigr]^a   
	= \int_0^{y(u)} \dd y'\, \frac{y'^a}{\dot{y}(u-\tau)}\,,
\end{align}
reading $\dot{y}(u-\tau)$ as a function of $y'$ from Eqs.~\eqref{eq:rdot_omegadot}, and supposing that $a > 4$, as is the case in practical computations. To 1.5PN relative order, we have
\begin{align}\label{eq:Ja0res}
	\mathcal{J}_{a,0} 
	= \frac{5 G m}{64c^3\nu}\,\frac{y^{a-4}}{a-4}\left[1 + \frac{a-4}{a-3} \left(\frac{743}{336}+\frac{11}{4}\nu\right)y -8\pi \frac{a-4}{2a-5} y^{3/2} + \calO\left(y^2\right)\right] \,.
\end{align}
It is interesting to observe that the tail effect in the flux directly influences the DC memory in~\eqref{eq:Ja0res}.

Last, but not least, we need to look at the interesting case of the tails-of-memory terms~\eqref{Uij4PN}. These terms can be treated as standard tail terms (with relative Newtonian accuracy), except for the ``genuine'' tail-of-memory given by the first line of~\eqref{Uij4PN}, namely
\begin{equation}\label{eq:ToMKij}
\mathcal{K}_{ij} = \frac{8G^2\dM}{7c^8}\int_0^{+\infty} \!\dd\rho\,  \dM_{a \langle i}^{(4)}(u-\rho) \int_0^{+\infty} \!\dd \tau\,  \dM_{j \rangle a}^{(4)}(u-\rho-\tau)  \ln\left(\frac{\tau}{\tau_0}\right)\,, 
\end{equation}
where $c \,\tau_0 = 2 r_0\,\de^\frac{1613}{270}$. We remind that this expression agrees with the tail-of-memory directly computed from the radiative quadrupole moment at infinity~\cite{F09,F11}. We first perform the tail-like integral over $\tau$. As we need to evaluate it at relative Newtonian order only, we can safely use the adiabatic approximation. Next we perform the integral over $\tau$, which is found to be a simple DC memory integral of the type~\eqref{eq:Ja0} with $a=13/2$. Hence we find
\begin{equation}\label{eq:ToMKijres}
\mathcal{K}_{ij} =
\frac{128\pi}{7}\frac{\nu^2 c^5}{G } \,\ell_{\langle i}\ell_{j \rangle} \int_0^{+\infty} \!\dd\rho\,\bigl[y(u-\rho)\bigr]^{13/2}
= \frac{4\pi}{7} m \nu \,c^2\,y^{5/2} \,\ell_{\langle i}\ell_{j \rangle}\,,
\end{equation}
where $\ell^i=L^i/|\bm{L}|$ is the constant unit vector associated with the Newtonian angular momentum and, thus, orthogonal to the orbital plane. Interestingly, this tail-of-memory result will give a contribution in the $(2,0)$ mode that exactly cancels the one coming from the 1.5PN corrections to the ordinary memory effect, obtained in~\eqref{eq:Ja0res}.

\section{Results}\label{sec:res}

Collecting all the pieces that were discussed in previous sections, and using notably the integration techniques developed in Sec~\ref{sec:toolbox}, we obtain our main results, namely the gravitational flux and quadrupole modes. All results displayed hereafter can be found in the ancillary file~\cite{AncFile}.

\subsection{Flux at 4PN order for generic orbits}\label{subsec:res_Fgen}

For generic orbits, and in the CoM frame, we split the gravitational energy flux~\eqref{Flux_Thorne} as
\begin{equation}\label{Flux_expl}
\mathcal{F} = \mathcal{F}_\text{can} + \mathcal{F}_\text{non-lin}\,.
\end{equation}
The first piece, $\mathcal{F}_\text{can}$, is the contribution due to the canonical mass and current moments, when they are fully expressed in terms of the compact binary parameters, \emph{i.e.}
\begin{equation}\label{Flux_can}
\mathcal{F}_\text{can} = \sum_{\ell \geq 2}\frac{G}{c^{2\ell+1}}\bigg[a_\ell\,\dM_L^{(\ell+1)}\dM_L^{(\ell+1)} + \frac{b_\ell}{c^2}\,\dS_L^{(\ell+1)}\dS_L^{(\ell+1)}\bigg] \,,
\end{equation}
where we recall that the numerical coefficients $a_\ell$ and $b_\ell$ are given in Eq.~\eqref{eq:aell_bell}. We further split $\mathcal{F}_\text{can}$ into a local-in-time (or instantaneous) part and a non-local one:
\begin{equation}\label{Flux_can2}
	\mathcal{F}_\text{can} = \mathcal{F}_\text{can}^\text{loc} + \mathcal{F}_\text{can}^\text{non-loc}\,.
\end{equation}
The local piece $\mathcal{F}_\text{can}^\text{loc}$ is too long to be displayed here but can be found in the ancillary file~\cite{AncFile}. Note the presence of the scales associated with the regularization processes, $r_0$ and $r_0'$, which is expected, as the canonical part of the flux is not a gauge invariant quantity. As for the non-local piece, $\mathcal{F}_\text{can}^\text{non-loc}$, it comes from two effects: (i) the direct contribution of the non-local part of the mass quadrupole moment, given by Eq.~(6.5) in~\cite{MQ4PN_renorm}; (ii) the tail term in the 4PN equations of motion, which contributes when time differentiating the 4PN mass quadrupole moment. [The latter tail term is given by the first line of Eq.~(4.4) in~\cite{BBFM17}, while the second line is included in the local part.] The non-local piece reads (at the required order, we are dealing with Newtonian quantities, so we can identify source and canonical moments):
\begin{align}
	\mathcal{F}_\text{can}^\text{non-loc} 
	& =
	\frac{48}{5}\frac{G^3\dM}{c^{13}} \,\dM_{ij}^{(3)}\Bigg\lbrace
	\frac{1}{7}\,\frac{\dd^3}{\dd u^3}\Bigg[\dM_{ik}(u) \int_0^{+\infty}\!\!\dd\tau \ln\left(\frac{c\tau}{2r_0}\right)\dM_{jk}^{(5)}(u-\tau)\Bigg]
	-\frac{m\nu}{15} x_{ik} \int_0^{+\infty}\!\!\dd\tau \ln\left(\frac{c\tau}{2r}\right)\dM_{jk}^{(8)}(u-\tau)\nn\\
	& \qquad\qquad\qquad\qquad
	- \frac{m\nu}{15}\big(3 x_kv_i +x_iv_k\big)\int_0^{+\infty}\!\!\dd\tau \ln\left(\frac{c\tau}{2r}\right)\dM_{jk}^{(7)}(u-\tau)\Bigg\rbrace+ \calO\left(\frac{1}{c^{14}}\right)\,.
\end{align}
Note the difference between the logarithmic kernel of the first integral, bearing $r_0$, and the two other ones, bearing $r$. Finally the second piece of the flux~\eqref{Flux_expl}, $\mathcal{F}_\text{non-lin}$, is given by all the non-linearities in the GW propagation discussed in Sec.~\ref{subsec:can2rad}. It contains all the double products between canonical moments and radiative moments, plus the square of the tail terms and the double product between the 1.5PN and 2.5PN corrections to the quadrupole. We write it as
\begin{align}
\mathcal{F}_\text{non-lin} =\
&
\frac{G}{5c^5}\bigg[
2\dM_{ij}^{(3)} \text{U}_{ij}^{\text{1.5PN}\,(1)}
+ 2\dM_{ij}^{(3)} \text{U}_{ij}^{\text{2.5PN}\,(1)}
+ 2\dM_{ij}^{(3)} \text{U}_{ij}^{\text{3PN}\,(1)} 
+ \text{U}_{ij}^{\text{1.5PN}\,(1)} \text{U}_{ij}^{\text{1.5PN}\,(1)}
+ \ldots \bigg]\nn\\
&\qquad
+ \frac{G}{189c^7}\bigg[
2\dM_{ijk}^{(4)} \text{U}_{ijk}^{\text{1.5PN}\,(1)}
+ 2\dM_{ijk}^{(4)} \text{U}_{ijk}^{\text{2.5PN}\,(1)} + \ldots \bigg] 
+ \ldots
+ \calO\left(\frac{1}{c^{14}}\right)\,,
\end{align}
where the ellipsis can be completed in a straightforward way from the results in Sec.~\ref{subsec:can2rad}.

\subsection{Flux at 4.5PN order for quasi-circular orbits}\label{subsec:res_Fcirc}

Introducing the unit separation $\bm{n}=\bm{x}/r$ between the particles, as well as the unit vector $\bm{\lambda} = \bm{\ell} \times \bm{n}$ such that $(\bm{n}, \bm{\lambda}, \bm{\ell})$ forms a direct orthonormal triad, we may write the relative velocity as $\bm{v} = r \omega \bm{\lambda} + \dot{r}\,\bm{n}$, where $\dot{r}$ is given by~\eqref{eq:rdot_omegadot}. Then, using~\eqref{keplerlaw}, we can express the 4PN flux on quasi-circular orbits as a function of the orbital frequency $\omega$, or equivalently the parameter $y$ defined by~\eqref{eq:y_def}. 

However we find that the 4PN flux parametrized by $y$ still contains the unphysical constant $b_0$, although the other arbitrary scales $r_0$ and $r'_0$ have properly disappeared. The reason is that, starting at the 4PN order, the frequency is modified due to the  propagation of tails in the wave zone. The half-phase $\psi$ of the reduced GW, \emph{i.e.} half the phase of the (2,2) mode, differs from the orbital phase $\phi$ (such that the orbital frequency is $\omega = \dot{\phi}$) by a logarithmic, tail-induced phase modulation, as
\begin{equation}\label{phasemod}
	\psi = \phi - \frac{2 G \dM \,\omega}{c^3} \ln\biggl(\frac{\omega}{\omega_0}\biggr)\,,
\end{equation}
where $\dM$ denotes the constant ADM mass, and $\omega_0$ is related to $b_0$ by $c\,\omega_0^{-1} = 4b_0 \,\de^{\gamma_\text{E}-11/12}$. We remind the reader that the constant $b_0$, which appears in the tail terms~\eqref{Uij15PN},~\eqref{Uij3PN} and~\eqref{Uij4PN}, is arbitrary and parametrizes the logarithmic deviation of light cones in harmonic coordinates from the flat cones, see \emph{e.g.}~(2.10)--(2.11) in~\cite{TB23}. 
%Another interpretation is that $b_0/c$ parametrizes a shift in the origin of the time coordinate. 
The constant $r_0$ is an IR scale introduced into the definition of the source multipole moments, see Eq.~(2.1) in~\cite{MHLMFB20}, while $r_0'$ is a UV scale associated with the regularization of the self-field of point particles.

This phase modulation was determined in~\cite{Wi93,BS93} and it has been repeatedly argued~\cite{BIWW96, ABIQ04} that it affects the waveform at the 4PN order. Now, the phase modulation~\eqref{phasemod} also shifts the GW half-frequency $\Omega = \dot{\psi}$ with respect to the orbital one $\omega = \dot{\phi}$. The GW half-frequency, directly measurable from the waveform at infinity, reads
\begin{equation}
	\Omega = \omega - \frac{2 G \dM \,\dot{\omega}}{c^3}\left[ \ln\biggl(\frac{\omega}{\omega_0}\biggr) + 1\right]\,.
\end{equation}
Using Eq.~\eqref{eq:rdot_omegadot} for the frequency chirp at the dominant order, it explicitly comes, replacing $\dM$ by $m$ at that order,
\begin{equation}\label{omOm}
	\Omega = \omega\,\Bigg\lbrace 1 - \frac{192}{5} \,\nu \left(\frac{G m \omega}{c^3}\right)^{8/3}\bigg[\ln\biggl(\frac{\omega}{\omega_0}\biggl) + 1 \bigg]  + \calO\left(\frac{1}{c^{10}}\right)\Bigg\rbrace\,,
\end{equation}
where we recall that $\nu = \frac{m_1m_2}{m^2}$ is the symmetric mass ratio. In the following, the results are expressed for quasi-circular orbits in terms of the measurable GW half-frequency $\Omega$ through the PN parameter
\begin{equation}\label{eq:x_def}
	x = \left(\frac{Gm\Omega}{c^3}\right)^{2/3}\,.
\end{equation}
Recalling also the definition~\eqref{eq:y_def} and posing $y_0 = (\frac{Gm\omega_0}{c^3})^{2/3}$ we thus have
\begin{equation}\label{xy}
	x = y\,\Bigg\lbrace 1 - \frac{192}{5} \,\nu \, y^4 \bigg[\ln\biggl(\frac{y}{y_0}\biggr) + \frac{2}{3} \Bigg]  + \mathcal{O}\bigl(y^5\bigr) \Biggr\rbrace\,,
\end{equation}
showing that indeed the GW half-frequency $\Omega$ differs from the orbital one $\omega$ at the 4PN order only, hence the fact that it did not affect previous computations such as in~\cite{BIWW96,ABIQ04}. Therefore, once the 4PN flux is obtained in terms of $y$, we replace $y$ in terms of $x$ using the inverse of Eq.~\eqref{xy} and reexpand to 4PN order. With this procedure, the constant~$b_0$ cancels out as expected. Finally, adding also the 4.5PN piece~\cite{MBF16}, the quasi-circular 4.5PN flux is
\begin{align}\label{Flux_x}
\mathcal{F} = \frac{32c^5}{5G}\nu^2 x^5 \Biggl\{
&
1 
+ \biggl(-\frac{1247}{336} - \frac{35}{12}\nu \biggr) x 
+ 4\pi x^{3/2}
+ \biggl(-\frac{44711}{9072} +\frac{9271}{504}\nu + \frac{65}{18} \nu^2\biggr) x^2 
+ \biggl(-\frac{8191}{672}-\frac{583}{24}\nu\biggr)\pi x^{5/2}
\nonumber\\
%%%%%%%%%%%%%%%%%%%%%%%%%%%%%%%%%%%%%%%%%%%%%%%%%%%%%%%%%%%%%%
& 
+ \Biggl[\frac{6643739519}{69854400}+ \frac{16}{3}\pi^2-\frac{1712}{105}\gamma_\text{E} - \frac{856}{105} \ln (16\,x) 
+ \biggl(-\frac{134543}{7776} + \frac{41}{48}\pi^2 \biggr)\nu 
- \frac{94403}{3024}\nu^2 
- \frac{775}{324}\nu^3 \Biggr] x^3 
\nonumber\\
%%%%%%%%%%%%%%%%%%%%%%%%%%%%%%%%%%%%%%%%%%%%%%%%%%%%%%%%%%%%%%
&
+ \biggl(-\frac{16285}{504} + \frac{214745}{1728}\nu +\frac{193385}{3024}\nu^2\biggr)\pi x^{7/2} 
\nonumber\\
%%%%%%%%%%%%%%%%%%%%%%%%%%%%%%%%%%%%%%%%%%%%%%%%%%%%%%%%%%%%%%
&
+ \Biggl[ -\frac{323105549467}{3178375200} + \frac{232597}{4410}\gamma_\text{E} - \frac{1369}{126} \pi^2 + \frac{39931}{294}\ln 2 - \frac{47385}{1568}\ln 3 + \frac{232597}{8820}\ln x   
\nonumber\\
%%%%%%%%%%%%%%%%%%%%%%%%%%%%%%%%%%%%%%%%%%%%%%%%%%%%%%%%%%%%%%
& \qquad
+ \biggl( -\frac{1452202403629}{1466942400} + \frac{41478}{245}\gamma_\text{E} - \frac{267127}{4608}\pi^2 + \frac{479062}{2205}\ln 2 + \frac{47385}{392}\ln 3  + \frac{20739}{245}\ln x \biggr)\nu
\nonumber\\
%%%%%%%%%%%%%%%%%%%%%%%%%%%%%%%%%%%%%%%%%%%%%%%%%%%%%%%%%%%%%%
& \qquad
+ \biggl( \frac{1607125}{6804} - \frac{3157}{384}\pi^2 \biggr)\nu^2 + \frac{6875}{504}\nu^3 + \frac{5}{6}\nu^4 \Biggr] x^4
\nonumber\\ 
%%%%%%%%%%%%%%%%%%%%%%%%%%%%%%%%%%%%%%%%%%%%%%%%%%%%%%%%%%%%%%
& 
+ \Biggl[ \frac{265978667519}{745113600} - \frac{6848}{105}\gamma_\text{E} - \frac{3424}{105} \ln (16 \,x)
+ \biggl( \frac{2062241}{22176} + \frac{41}{12}\pi^2 \biggr)\nu
\nonumber\\ 
%%%%%%%%%%%%%%%%%%%%%%%%%%%%%%%%%%%%%%%%%%%%%%%%%%%%%%%%%%%%%%
& \qquad
- \frac{133112905}{290304}\nu^2 - \frac{3719141}{38016}\nu^3 \Biggr] \pi x^{9/2} 
+ \mathcal{O}\bigl(x^5\bigr) \Biggr\}\,.
\end{align}
A significant check is to observe that the leading-order terms in the test-mass limit $\nu\to 0$ perfectly agree with the results of linear black-hole perturbation theory~\cite{Sasa94,TSasa94,TTS96,Fuj14PN,Fuj22PN}. Note that the flux has been confirmed by different groups up to 2PN order~\cite{WW96,LMRY19}, and that the 4.5PN piece is in agreement with the independent work~\cite{Messina:2017yjg}. All other terms are new with the present paper. We have also explicitly verified that, at the 4PN order, Eq.~\eqref{Flux_x} can be recovered from the gravitational modes given by Eq.~\eqref{h22} below and in~\cite{Henry:2022ccf}, using
\begin{equation}
\mathcal{F} = \frac{c^3}{16\pi G}\sum_{\ell \geqslant 0} \sum_{\text{m} = - \ell}^\ell\,\big\vert \dot{h}_{\ell \text{m}}\big\vert^2\,.
\end{equation}
As usual, the contributions due to the absorption by the black-hole horizons should be added separately from the post-Newtonian result~\eqref{Flux_x}; see~\cite{PS95,TMT97,Alvi01,Porto:2007qi,Chatz12,Saketh:2022xjb}.

In the companion Letter~\cite{BFHLT_Letter} we use the quasi-circular flux $\mathcal{F}(x)$ to derive the gravitational frequency chirp and phase at the 4.5PN order. Namely, we apply the energy flux-balance law 
\begin{equation}\label{flux-balance}
\dot{E}(x)=-\mathcal{F}(x)\,,
\end{equation}
where $x$ is directly linked to the observed GW half-frequency through the definition~\eqref{eq:x_def}, and is related to the orbital frequency \emph{via} the relation~\eqref{xy}. 
Here, $E(x)$ denotes the (Noetherian) binding energy of the compact binary on the quasi-circular orbit, as calculated from the 4PN equations of motion, see \emph{e.g.}~\cite{BBFM17}. The fact that the left-hand side of the balance equation, which concerns the motion, should be expressed in terms of the same observed GW half-frequency $x$ as the right-hand side, which concerns the radiation, and not, for instance, in terms of the ``orbital'' frequency $y$, is worth an explanation: suppose that the compact binary system is actually a binary pulsar system. Hence, in addition to the gravitational waves generated by the orbital motion, the pulsar emits electromagnetic (radio) waves, also received by the far away observer. Now the observer at infinity can measure the orbital frequency of the system from the instants of arrival of the radio pulses --- this is the standard analysis of binary pulsars. Such frequency should be the one to be inserted in the left side of the balance equation. However, far from the system, the space-time curvature $\mathcal{R}^{-1}\sim\sqrt{\dM/r^3}$ tends to zero, and therefore the geometric optics or WKB approximation applies for both the EM and gravitational waves. Thus the EM and gravitational waves follow the same geodesic, independently of their frequency, and in particular are subject to the same tail-induced phase modulation~\eqref{phasemod}. We conclude that the distant observer measures the same frequency $x$ from the EM radio pulses and from the gravitational wave, and this is that frequency that he inserts into both sides of the flux-balance law~\eqref{flux-balance}.\footnote{See~\cite{Detweiler:2005kq} for a similar argument in the context of self-forces, using an observer sitting on the particle and equipped with a flashlight.}

\subsection{Gravitational wave modes \boldmath $(2,2)$ \unboldmath and \boldmath $(2,0)$ \unboldmath}\label{subsec:res_modes}

Next, we can extract the $(\ell,\text{m}) = (2,2)$ and $(2,0)$ physical modes from the newly computed 4PN mass quadrupole radiative moment. Projecting the asympotic metric~\eqref{eq:hTT} onto the basis of polarizations $\{+,\times\}$ and the usual basis of spin-weighted spherical harmonics, $Y_{-2}^{\ell \text{m}}$ (following the conventions of~\cite{BFIS08,FMBI12}), one can define the observable gravitational modes $h_{\ell \text{m}}$ as
\begin{equation}
h_+ - \di h_\times
= \sum_{\ell = 2}^{+\infty}\sum_{\text{m}=-\ell}^{\ell}h_{\ell \text{m}} \,Y_{-2}^{\ell \text{m}}\,.
\end{equation}
For the sake of simplicity, we single out the (observable) GW half-phase $\psi$, and rescale the modes by the dominant contribution, defining $H_{\ell \text{m}}$ as
\begin{equation}
h_{\ell \text{m}}
= \frac{8 G m \nu x}{c^2 R} \,\sqrt{\frac{\pi}{5}} \,H_{\ell \text{m}} \, \de^{-\di \text{m} \psi}\,.
\end{equation}
We have $H_{\ell,-\text{m}} = (-)^\ell \,\overline{H}_{\ell \text{m}}$, where the overbar denotes the complex conjugate. Using the PN-MPM framework, the current state-of-the-art is the 3.5PN accuracy for all the modes~\cite{FMBI12,Henry:2022dzx,Henry:2022ccf}. From the present completion of the 4PN mass quadrupole $\text{U}_{ij}$, one can extract the 4PN correction to the $(\ell,\text{m}) = (2,2)$ mode as
\begin{equation}
H_{22} = -\frac{\de^{2\di \psi}}{2c^2m\nu x} \overline{\mathfrak{m}}^i \overline{\mathfrak{m}}^j \,\text{U}_{ij}\,,
\end{equation}
where $\bm{\mathfrak{m}}=(\bm{n}+ \di \bm{\lambda})/\sqrt{2}$, with $\overline{\bm{\mathfrak{m}}}$ its complex conjugate, and the radiative moment $\text{U}_{ij}$ is evaluated at the retarded time $u$. Explicitly, it comes
\begin{align} \label{h22}
	H_{22} &= 1 + \biggl(-\frac{107}{42}+\frac{55}{42}\nu\biggr) x 
	+2 \pi x^{3/2} 
	+ \biggl(-\frac{2173}{1512}-\frac{1069}{216}\nu+\frac{2047}{1512}\nu^2\biggr)x^2
	+  \left[-\frac{107 \pi }{21} +\left(\frac{34 \pi}{21}-24 \,\di\right)\nu\right]x^{5/2} \nonumber \\
	%%%%%%%%%%%%%%%%%%%%%%%%%%%%%%%%%%%%%%%%%%%%%%%%%%%%%%%%%%%%%%%%%%%%%%%%%%%%%%%
	&\qquad + \Biggl[\frac{27027409}{646800}-\frac{856}{105}\,\gamma_\text{E} +\frac{428\,\di\,\pi }{105}+\frac{2 \pi ^2}{3} + \biggl(-\frac{278185}{33264}+\frac{41 \pi^2}{96}\biggr) \nu -\frac{20261}{2772}\nu^2+\frac{114635}{99792}\nu^3-\frac{428}{105} \ln (16 x)\Biggr]x^3
	\nonumber \\
	%%%%%%%%%%%%%%%%%%%%%%%%%%%%%%%%%%%%%%%%%%%%%%%%%%%%%%%%%%%%%%%%%%%%%%%%%%%%%%%
	&\qquad +  \Biggl[ -\frac{2173\pi}{756} + \biggl(
	-\frac{2495\pi}{378}+\frac{14333\,\di}{162} \biggr)\nu + \biggl(
	\frac{40\pi}{27}-\frac{4066\,\di}{945} \biggr)\nu^2 \Biggr] x^{7/2} 
	\nonumber \\
	%%%%%%%%%%%%%%%%%%%%%%%%%%%%%%%%%%%%%%%%%%%%%%%%%%%%%%%%%%%%%%%%%%%%%%%%%%%%%%%
	&\qquad + \Biggl[- \frac{846557506853}{12713500800} + \frac{45796}{2205}\gamma_\text{E} - \frac{22898}{2205}\di \pi - \frac{107}{63}\pi^2 + \frac{22898}{2205}\ln(16x) \nonumber \\
	%%%%%%%%%%%%%%%%%%%%%%%%%%%%%%%%%%%%%%%%%%%%%%%%%%%%%%%%%%%%%%%%%%%%%%%%%%%%%%%
	&\qquad\qquad  +\biggl(- \frac{336005827477}{4237833600} + \frac{15284}{441}\gamma_\text{E} - \frac{219314}{2205}\di \pi - \frac{9755}{32256}\pi^2 + \frac{7642}{441}\ln(16x)\biggr)\nu \nonumber \\
	%%%%%%%%%%%%%%%%%%%%%%%%%%%%%%%%%%%%%%%%%%%%%%%%%%%%%%%%%%%%%%%%%%%%%%%%%%%%%%%
	&\qquad\qquad +\biggl( \frac{256450291}{7413120} - \frac{1025}{1008}\pi^2 \biggr)\nu^2 - \frac{81579187}{15567552}\nu^3 + \frac{26251249}{31135104}\nu^4 \Biggr] x^4 +
	\mathcal{O}\bigl(x^{9/2}\bigr) \,.
\end{align}
In the test-mass limit $\nu\to 0$, this new result is also nicely in agreement with the prediction of linear black-hole perturbation theory~\cite{FI10}. 

Note that the two unphysical scales $r_0$ and $b_0$ appearing in the intermediate calculations are absent from the final results. More precisely, these constant appear in the expressions of the source moments and in the relations between radiative and source moments, notably in~\eqref{Uij4PN}. Although this is expected, the constants in both contributions cancel each other and this constitutes a highly non-trivial check of the MPM-algorithm and the robustness of the PN integration of the source moments.

Finally, we have also extracted the zero-frequency quadrupole mode $(\ell,\text{m})=(2,0)$ at 4PN order:
\begin{equation}
	H_{20} = -\sqrt{\frac{3}{8}}\,\frac{\ell^{\langle i}\ell^{j\rangle}}{c^2m\nu x}\,\text{U}_{ij}\,.
\end{equation}
As discussed in Sec.~\ref{subsec:memoryInt}, this mode arises from integration of the non-linear DC memory terms over the past history of the system, assuming a model for the quasi-circular evolution of the orbit in the past. Since the integration increases the effect by the inverse of the 2.5PN order, with the present 4PN formalism we are able to control the $(2,0)$ mode only with relative 1.5PN precision. We find 
\begin{align}\label{h20}
H_{20} &= - \frac{5}{14\sqrt{6}}\,\biggl[1 + \biggl(-\frac{4075}{4032}+\frac{67}{48}\nu\biggr) x + \mathcal{O}\bigl(x^2\bigr)\biggr] \,.
\end{align}
Notice that the 1.5PN term of this mode vanishes. This is due to the fact that the 1.5PN correction in the model of evolution of the quasi-circular orbit in the past, which results in the 1.5PN term in Eq.~\eqref{eq:Ja0res}, exactly cancels the 1.5PN direct contribution of the ``tail-of-memory'' at 4PN order, and given by~\eqref{eq:ToMKijres}. That is,
\begin{equation}
H_{20}^\text{ToM} = - H_{20}^\text{mem,1.5PN} = - \frac{2\sqrt{2}}{7\sqrt{3}} \pi \,x^{3/2}\,.
\end{equation}
The result~\eqref{h20} is in full agreement with Eq.~(4.3a) of~\cite{F09}, obtained from the general expression of non-linear memory terms in terms of radiative moments. Indeed, recall that the tail-of-memory integral~\eqref{eq:ToMKij} at 4PN order can be simply obtained from the leading memory integral~\eqref{Uij25PN} at 2.5PN order by replacing the canonical moment by the corresponding radiative moment including the tail effect~\eqref{Uij15PN} at relative 1.5PN order, see Eq.~\eqref{memToM}. This confirms that the tail-of-memory is adequately taken into account in the computation of the memory using radiative moments defined at future null infinity~\cite{F09,F11}.

\acknowledgments

It is a great pleasure to thank Laura Bernard, Alessandra Buonanno, Bala Iyer, Tanguy Marchand, Sylvain Marsat, Rafael Porto and Adam Pound for enlightening exchanges during the project. F.L. is grateful to the Institut d'Astrophysique de Paris for its hospitality during this project. He received funding from the European Research Council (ERC) under the European Union's Horizon 2020 research and innovation program (grant agreement No 817791). G.F. thanks IIT Madras for a Visiting Faculty Fellow position under the IoE program during the completion of this work.

\appendix

\section{Test of the boosted Schwarzchild solution}\label{app:BSS}

Among all tests that can be performed to check the expression of the mass quadrupole moment, one of the simplest is the boosted Schwarzschild limit. Despite its apparent simplicity, it is quite efficient and was crucially used to fix a remaining ambiguity constant in an early computation of the flux at 3PN order~\cite{BDI04zeta}. The principle is quite transparent: if we remove one of the two black holes, then our system reduces to a single Schwarzchild black hole of mass $m_1$, boosted at a (constant) speed $\bm{v}_1$. The multipole moments of a boosted Schwarzschild solution (BSS) are straightforward to derive, and have been previously determined at 3PN order in~\cite{BDI04zeta}. Extending this work at 4PN order, we find that the quadrupole moment of a BBS of mass $\dM$, boosted with a velocity $\bm{V}$, reads
\begin{align}\label{Qij_BSS}
\dI_{ij}^\mathrm{BSS} =\ &
\dM\,t^2 \,V^{\langle i}V^{j\rangle}\left[
1+\frac{9}{14}\,\frac{V^2}{c^2}
+\frac{83}{168}\,\frac{V^4}{c^4}
+\frac{507}{1232}\,\frac{V^6}{c^6} 
+\frac{45923}{128128}\,\frac{V^8}{c^8} \right]\nn\\
& + 
\frac{4}{7}\,\frac{G^2\,\dM^3}{c^6}\,V^{\langle i}V^{j\rangle} 
+ \frac{10}{21}\,\frac{G^2\,\dM^3}{c^8}\,V^2\,V^{\langle i}V^{j\rangle}
+ \mathcal{O}\left(\frac{1}{c^{10}}\right)\,.
\end{align}
On the other hand, taking the BSS limit $(m_2,\bm{v}_2) \to (0,\bm{0})$ of the renormalized mass quadrupole moment defined in~\cite{MQ4PN_renorm} (on generic orbits, out of the CoM frame), it comes
\begin{align}\label{Iij_BSS}
\lim_\text{BSS}\dI_{ij} = &
m_1\,t^2 \,v_1^{\langle i}v_1^{j\rangle}\left[
1+\frac{9}{14}\,\frac{v_1^2}{c^2}
+\frac{83}{168}\,\frac{v_1^4}{c^4}
+\frac{507}{1232}\,\frac{v_1^6}{c^6}
+\frac{45923}{128128}\,\frac{v_1^8}{c^8}\right]\nn\\
& +
\frac{4}{7}\frac{G^2\,m_1^3}{c^6}\,v_1^{\langle i}v_1^{j\rangle} 
+ \frac{10}{21}\,\frac{G^2\,m_1^3}{c^8}\,v_1^2\,v_1^{\langle i}v_1^{j\rangle} 
+ \mathcal{O}\left(\frac{1}{c^{10}}\right)\,.
\end{align}
Both expressions coincide under the identification $(m_1,\bm{v}_1) = (\dM,\bm{V})$, therefore this test is conclusive.

\section{Corrections to the metric due to infrared commutators}
\label{app:comm_metric}

This appendix displays the formal contributions of the infrared commutators to the gothic metric perturbation $h^{\mu\nu} = \sqrt{-g}g^{\mu\nu} - \eta^{\mu\nu}$, up to 4PN order, as discussed in Sec.~\ref{sec:commutators}. Those contributions are to be added to Eqs.~(A.2) of the work~\cite{MHLMFB20} in order to obtain the full metric at 4PN, and the potentials entering it are the one defined by Eqs.~(A.4) of~\cite{MHLMFB20}. It thus comes, at the required order [the commutators being defined by~\eqref{comm_F_def}]
\begin{subequations}\label{eq:app_comm_hmunu}
	\begin{align}
		\Delta h^{00} = & \nonumber
		- \frac{2(d-1)^2}{(d-2)^2c^4}\,\comm\big\lbrace V^2\big\rbrace\\
		& \nonumber
		+ \frac{1}{c^6} \Bigg[
		\frac{8(d-1)^2(d-3)}{(d-2)^3}\,\comm\big\lbrace KV\big\rbrace
		+\frac{8(d-3)}{d-2}\,\comm\big\lbrace V_iV_i\big\rbrace- \frac{4(d-1)}{d-2}\,\comm\big\lbrace V\hat{W}\big\rbrace\\
		& \qquad\quad\nonumber
		- \frac{4(d-1)^3}{(d-2)^3}\bigg(\frac{1}{3}\comm\big\lbrace V^3\big\rbrace + V\,\comm\big\lbrace V^2\big\rbrace+\comm\big\lbrace V\comm\big\lbrace V^2\big\rbrace\big\rbrace\bigg)
		\Bigg]\\
		& \nonumber
		+ \frac{1}{c^8}\Bigg[
		\frac{2(d-1)^3(4-3d)}{(d-2)^4}\bigg(\frac{V}{3}\,\comm\big\lbrace V^3\big\rbrace+ V\,\comm\big\lbrace V\,\comm\big\lbrace V^2\big\rbrace\big\rbrace  + \frac{1}{4}\left(\comm\big\lbrace V^2\big\rbrace\right)^2 - \frac{2(d-1)\,V^2}{4-3d}\,\comm\big\lbrace V^2\big\rbrace\bigg)\\
		& \qquad\quad \nonumber
		- \frac{4(d-1)^2(d-3)(4-3d)}{(d-2)^4}\bigg(V\,\comm\big\lbrace KV\big\rbrace + \frac{K}{2}\,\comm\big\lbrace V^2\big\rbrace\bigg)
		- \frac{4(d-3)(4-3d)}{(d-2)^2}\,V\,\comm\big\lbrace V_iV_i\big\rbrace\\
		& \qquad\quad
		+ \frac{8(d-1)(d-4)}{(d-2)^2}\,V_i\,\comm\big\lbrace VV_i\big\rbrace
		+ \frac{2(d-1)(4-3d)}{(d-2)^2}\,V\,\comm\big\lbrace V\hat{W}\big\rbrace
		- \frac{4(d-1)^2}{(d-2)^2}\,\hat{W}\,\comm\big\lbrace V^2\big\rbrace
		\Bigg]
		+ \calO\left(\frac{1}{c^{10}}\right)\,,\\
		%======================
		\Delta h^{0i} = & \nonumber
		\,\frac{4(d-1)}{(d-2)c^5}\,\comm\big\lbrace VV_i\big\rbrace\\
		& \nonumber
		+ \frac{1}{c^7} \Bigg[
		\frac{4(d-1)^2}{(d-2)^2}\bigg(\comm\big\lbrace V^2V_i\big\rbrace+V_i\,\comm\big\lbrace V^2\big\rbrace + V\,\comm\big\lbrace VV_i\big\rbrace + \comm\big\lbrace V_i\,\comm\big\lbrace V^2\big\rbrace\big\rbrace + \comm\big\lbrace V\,\comm\big\lbrace VV_i\big\rbrace\big\rbrace\bigg)\\
		& \qquad\quad
		- \frac{8(d-1)(d-3)}{(d-2)^2}\,\comm\big\lbrace KV_i\big\rbrace
		+ 8\,\comm\big\lbrace \hat{W}V_i\big\rbrace
		- 8 \,\comm\big\lbrace \hat{W}_{ik}V_k\big\rbrace
		+ \frac{8(d-1)}{d-2}\,\comm\big\lbrace V\hat{R}_i\big\rbrace\Bigg]
		+ \calO\left(\frac{1}{c^9}\right)\,,\\
		%======================
		\Delta h^{ij} = & \nonumber
		- \frac{8(d-1)}{(d-2)c^8}\bigg(V_i\,\comm\big\lbrace VV_j\big\rbrace+V_j\,\comm\big\lbrace VV_i\big\rbrace\bigg)\\
		& \nonumber
		+\frac{\delta_{ij}}{c^8} \Bigg[
		\frac{4(d-1)^2(d-3)}{(d-2)^3}\bigg(V\,\comm\big\lbrace KV\big\rbrace +\frac{K}{2}\,\comm\big\lbrace V^2\big\rbrace\bigg)
		- \frac{2(d-1)^3}{(d-2)^3}\bigg(\frac{V}{3}\,\comm\big\lbrace V^3\big\rbrace +V\,\comm\big\lbrace V\,\comm\big\lbrace V^2\big\rbrace\big\rbrace+ \frac{1}{4}\left(\comm\big\lbrace V^2\big\rbrace\right)^2\bigg)\\
		& \qquad\quad
		+ \frac{4(d-3)}{d-2}\,V\,\comm\big\lbrace V_kV_k\big\rbrace
		+ \frac{8(d-1)}{d-2}\,V_k\,\comm\big\lbrace VV_k\big\rbrace
		- \frac{2(d-1)}{d-2}\,V\,\comm\big\lbrace V\hat{W}\big\rbrace
		\Bigg]
		+ \calO\left(\frac{1}{c^{10}}\right)\,.
	\end{align}
\end{subequations}

\bibliography{ListeRef_flux4PN.bib}

\end{document}